\documentclass[aps,pre,twocolumn,showpacs,superscriptaddress,groupedaddress]{revtex4-1} 

\usepackage{graphicx}        
\usepackage{dcolumn}        
\usepackage{bm}                
\usepackage{amssymb}
\usepackage{amsmath}        
\usepackage{color}
\definecolor{korr_26Apr}{rgb}{0,0,0} 
\definecolor{red}{rgb}{1,0,0} 

\renewcommand{\figurename}{Fig.}

\def \d{\mathrm{d}}

\begin{document}

\widetext



\title{Analytical model for flux saturation in sediment transport}
\author{Thomas P\"ahtz$^{1,2}$, Eric J. R. Parteli$^3$, Jasper F. Kok$^4$ and Hans J. Herrmann$^{5,6}$}
\affiliation{1.~Ocean College, Zhejiang University, 310058 Hangzhou, China. \\
2.~State Key Laboratory of Satellite Ocean Environment Dynamics, Second Institute of Oceanography, 310012 Hangzhou, China. \\
3.~${\mbox{Institute for Multiscale Simulation, Universit\"at Erlangen-N\"urnberg, N\"agelsbachstra{\ss}e~49b, 91052 Erlangen, Germany.}}$ \\
4.~${\mbox{Department of Atmospheric and Oceanic Sciences, University of California, Los Angeles, California 90095, USA.}}$ \\
5.~Departamento de F\'isica, Universidade Federal do Cear\'a, 60451-970 Fortaleza, Cear\'a, Brazil. \\
6.~Computational Physics, IfB, ETH Z\"urich, Schafmattstra{\ss}e~6, 8093 Z\"urich, Switzerland.}

\begin{abstract}
The transport of sediment by a fluid along the surface is responsible for dune formation, dust entrainment and for a rich diversity of patterns on the bottom of oceans, rivers, and planetary surfaces. Most previous models of sediment transport have focused on the equilibrium (or saturated) particle flux. However, the morphodynamics of sediment landscapes emerging due to surface transport of sediment is controlled by situations out-of-equilibrium. In particular, it is controlled by the saturation length characterizing the distance it takes for the particle flux to reach a new equilibrium after a change in flow conditions. The saturation of mass density of particles entrained into transport and the relaxation of particle and fluid velocities constitute the main relevant relaxation mechanisms leading to saturation of the sediment flux. Here we present a theoretical model for sediment transport which, for the first time, accounts for both these relaxation mechanisms and for the different types of sediment entrainment prevailing under different environmental conditions. Our analytical treatment allows us to derive a closed expression for the saturation length of sediment flux, which is general and can thus be applied under different physical conditions.

\end{abstract}


\maketitle

\section{Introduction}\label{Introduction}

When a sediment bed is exposed to a fluid flow, particles can be entrained and transported by different mechanisms. The transport regime depends primarily on the inertial characteristics of the particles and the fluid. Sufficiently light particles are transported as suspended load in which their weight is supported by the turbulence of the fluid. In contrast, particles which are sufficiently heavy are transported along the surface \cite{Bagnold41,VanRijn93}. This type of transport incorporates two main transport modes, namely: {\em{saltation}}, which consists of sediment grains jumping downstream close to the ground at nearly ballistic trajectories, and {\em{creep}}, which consists of particles rolling and sliding along the sediment bed. Sediment transport along the surface is responsible for a wide range of geophysical phenomena, including surface erosion, dust aerosol emission, and the formation and migration of dunes \cite{Bagnold41,VanRijn93,GreeleyIversen85,Shao08,Kroyetal02a,Koketal12}. Therefore, the quantitative understanding of sediment transport may improve our understanding of river beds evolution \cite{VanRijn93}, the emission of atmospheric dust \cite{Shao08,Koketal12} and the dynamics of planetary sand landscapes \cite{GreeleyIversen85,Bourkeetal10,Koketal12}.

Once sediment transport begins, the fluid loses momentum to accelerate the particles as a consequence of Newton's second law (the transport-flow feedback, e.g. \cite{UngarHaff87,Almeidaetal07,Almeidaetal08,Paehtzetal12a}). Therefore, the sediment flux, $Q$, which is the average momentum of grains transported per unit soil area, is limited by an equilibrium value, the saturated flux, $Q_s$. Although previous studies focused on this equilibrium flux (e.g. \cite{MeyerPeterMuller48,Einstein50,Bagnold66,Sorensen91,AbrahamsGao06,Laemmeletal12}), the dynamics of sediment landscapes is controlled by situations {\em{out-of-equilibrium}}. In particular, the sediment flux needs a spatial lag --- the so-called saturation length, $L_s$ --- to adapt to a change in flow conditions \cite{Sauermannetal01,ClaudinAndreotti06,Andreottietal10,Fourriereetal10}. This saturation length introduces the main relevant length-scale in the dynamics of sediment landscapes under water and on the surface of planetary bodies. For instance, the saturation length controls the minimal size of crescent-shaped (barchan) dunes moving on top of bedrock, as well as the wavelength of the smallest dunes (the so-called ``elementary dunes'') emerging on top of a sediment bed \cite{ClaudinAndreotti06,Fourriereetal10}. Although important insights were gained recently from experimental studies \cite{Hersenetal02,ClaudinAndreotti06,FranklinCharru11}, the physics behind the saturation length, and thus the dependence of $L_s$ on flow and sediment attributes, is still insufficiently understood. 

One of the most important deficiencies in our understanding of the dependence of $L_s$ on flow and sediment conditions is that it remains uncertain which mechanisms are most important in determining the saturation of the sediment mass flux. On the one hand, it has been suggested that the acceleration of transported particles due to fluid drag is the dominant relaxation mechanism \cite{Hersenetal02,ClaudinAndreotti06,Andreottietal10,Fourriereetal10}. This model neglects the entrainment of sediment bed particles due to fluid lift, as well as the entrainment of sediment bed particles and the deceleration of transported particles due to collisions of transported particles with the sediment bed (grain-bed collisions). On the other hand, the entrainment of sediment bed particles by fluid lift and grain-bed collisions has also been proposed to be the dominant relaxation mechanisms \cite{Sauermannetal01,Charru06}. However, these models neglect momentum changes of transported particles, which is exactly the opposite situation of the models in Refs.~\cite{Hersenetal02,ClaudinAndreotti06,Andreottietal10,Fourriereetal10}. Moreover, to our knowledge, all previous models neglected a further relaxation mechanism of the sediment flux, namely, the relaxation of the fluid speed in the transport layer ($U$) due to the saturation of the transport-flow feedback \cite{MaZheng11}.

To address this situation and develop an accurate expression for $L_s$ that can be used in future studies, this paper presents a model for flux saturation in sediment transport which, for the first time, accounts for {\em{all}} aforementioned mechanisms for the saturation of sediment flux. In particular, our theoretical model accounts for the coupling between the entrainment of sediment bed particles due to fluid lift and grain-bed collisions, the acceleration and deceleration of transported particles due to fluid forces and grain-bed collisions, and the saturation of $U$ due to the saturation of the transport-flow feedback. Our analytical model allows us to derive a closed expression for $L_s$ which can be applied to different physical environments. Our model suggests that grain-bed collisions, which have been neglected in all previous studies, have an important influence on the saturation length, $L_s$. Moreover, our model suggests that the relaxation of $U$ plays an important role for sediment transport in dilute fluids (aeolian transport), whereas it plays a negligible role for sediment transport in dense fluids (subaqueous transport).

In a recent Letter (see Ref.~\cite{Paehtzetal13}), we presented our equation for $L_s$ and showed that it is consistent with measurements of $L_s$ in both subaqueous and aeolian sediment transport regimes over at least five orders of magnitude in the ratio between fluid and particle density. In the present paper, we derive the analytical model presented in Ref.~\cite{Paehtzetal13} in more detail and study the properties of the equations governing the behavior of the saturation length in both transport regimes. Since Ref.~\cite{Paehtzetal13} includes a detailed comparison of our model against measurements, no model comparisons against measurements are included here.


This paper is organized as follows. Sections \ref{Derivation} and \ref{Obtaining_Ls} discuss the analytical treatment of flux saturation. In the former Section, we derive the mass and momentum conservation equations for the layer of sediments in transport, as well as the differential equation of the sediment flux in terms of the mass density and average velocity of the transported particles. These equations allow us to obtain a mathematical expression for the saturation length of sediment transport, which is presented in Section \ref{Obtaining_Ls}. This Section also discusses how to determine the quantities appearing in the saturation length equation, which encode the attributes of sediment and flow, as well as the characteristics of sediment entrainment and particle-fluid interactions. In Section \ref{Discussion} we use our theoretical expression to perform a study of the saturation length as a function of the relevant physical quantities controlling saturation of sediment flux. Conclusions are presented in Section \ref{Conclusions}.

\section{Flux saturation in sediment transport}\label{Derivation}

The downstream evolution of the sediment flux, $Q$, towards its equilibrium value, $Q_s$, can be described by the following equation \cite{Andreottietal10}, which is identical to Eq.~(1) of Ref.~\cite{Paehtzetal13},
\begin{eqnarray}
\Gamma(Q) = \frac{\d Q}{\d x}\approx\frac{Q_s-Q}{L_s}, \label{Lsdef1}
\label{Lsdef}
\end{eqnarray}
which is valid in the regime where $Q$ is close to saturation ($|1-Q/Q_s|\ll1$). The length-scale $L_s$, the saturation length, characterizes the response of the sediment flux due to a small change in flow conditions around equilibrium. Since $\Gamma(Q_s)=0$, $L_s$ can be written as the negative inverse first-order Taylor coefficient of $\Gamma(Q)$,
\begin{eqnarray}
 L_s=-\left(\frac{\d\Gamma}{\d Q}\right)_{Q=Q_s}^{-1}. \label{Lsdef2}
\end{eqnarray} 

In this Section, we derive the equations that describe the downstream evolution of the sediment mass flux, $Q$, towards its equilibrium value, $Q_{\mathrm{s}}$, in sediment transport under turbulent boundary layer flow. 

The mass flux $Q$ is defined as $Q = MV$, where $M$ is the average transported mass per unit soil area and $V$ is the average particle velocity. Therefore, the saturation of $Q$ is dictated by the mechanisms governing the relaxation of $M$ and $V$ towards their saturated values, $M_s$ and $V_s$, respectively. The quantitative description of the saturation processes of $M$ and $V$ requires incorporation of all relevant forces acting on the sediment particles in transport, namely drag, gravity, buoyancy, collision forces between particles in transport (``mid-fluid collisions'') and friction due to collisions between particles and the bed. Indeed, Moraga et al. \cite{Moragaetal99} found experimentally that lift forces due to shear flow acting on a particle surrounded by fluid --- which have often been assumed to be significant during transport (e.g. \cite{VanRijn93}) --- are approximately an order of magnitude smaller than the drag force and can be, thus, neglected in our calculations. On the other hand, the so-called added mass force exerted by accelerated or decelerated particles to dislodge the fluid as they move through it leads to enhanced inertia of the particles in transport. This added mass effect plays a relevant role for the motion of the particles \cite{NinoGarcia98a}, and thus we also take it into account. Our analytical treatment applies to situations where the fluid velocity is not too high such that only transport through saltation or creep (the main transport modes of particles along the surface \cite{Bagnold41,Koketal12}) is considered. Transport through suspension or dense transport regimes, such as sheet flow \cite{Gao08}, are, thus, not considered.

In Section \ref{definitions} we first present the definitions and notations used in our study. Afterwards in Section \ref{mass_and_momentum}, we present the local conservation equations, from which we obtain the saturation equations, presented in Section \ref{sec:saturation}.

\subsection{Definitions and Notations}\label{definitions}
We use a three-dimensional coordinate system $(x,y,z)$, where $x$ denotes the direction of fluid motion, $y$ is the lateral direction and $z$ is the vertical direction. The top of the sediment bed, which corresponds to the height at which the local particle concentration equals approximately $50\%$ of the particle concentration deep within the bed \cite{Duranetal12}, is located at the vertical position $z = h_o(x,y)$. Here we use the approximation that the slopes of bedforms are usually very small ($\partial h_o/\partial x\approx0$). Moreover, since the time-scale of the relaxation of the sediment flux due to changes in the flow is typically much smaller than the time-scale of the evolution of bedforms (dunes and ripples) \cite{Fourriereetal10} ($T_\mathrm{fl}\ll T_\mathrm{bed}$), we can adopt the approximation that the transport over the sediment landscape is in the steady-state, i.e. $\partial/\partial t=0$, where $t$ denotes time. Furthermore, since our description relates to the saturation of the mass flux $Q$ due to changes in the downstream direction, we consider a laterally invariant sediment bed ($\partial/\partial y=0$).

We consider a certain microscopic configuration of $N$ particles (including the limit $N\rightarrow\infty$) labeled by an upper index $n$ whose centers of mass are located at $\mathbf{x}^n$. Each particle has a mass $m^n$, a velocity $\mathbf{v}^n$, and is subjected to a force $\mathbf{F}^n$ resulting in an acceleration $\mathbf{a}^n=\mathbf{F}^n/m^n$. These forces include both external body forces ($\mathbf{F}^{\mathrm{ex}\,n}=m^n\mathbf{a}^{\mathrm{ex}\,n}$) and interparticle contact forces. In general these forces are non-conservative. The interparticle contact forces occur for all pairs of contacting particles. We therefore denote them by $\mathbf{F}^{mn}=-\mathbf{F}^{nm}$, which is the contact force applied by the particle with the number $n$ on the particle with the number $m$. We note that $\mathbf{F}^{mn}=0$ if these particles are not in contact, and we define $\mathbf{F}^{mm}=0$ (no self-interaction). Hence, the total acceleration of particle $n$ can be written as,
\begin{eqnarray}
 \mathbf{a}^n=\frac{1}{m^n}\sum_m\mathbf{F}^{nm}+\mathbf{a}^{\mathrm{ex}\,n}.
\end{eqnarray}

We define $f(\mathbf{x},\mathbf{v},m,t)$, the density of a certain microscopic configuration of particles at time $t$, as
\begin{equation}
 f(\mathbf{x},\mathbf{v},m,t)=\sum_n\delta(\mathbf{x}-\mathbf{x}^n(t))\delta(\mathbf{v}-\mathbf{v}^n(t))\delta(m-m^n). \label{def_f2}
\end{equation}
It describes the number of particles, $\d N$, with positions, velocities, and masses in infinitesimal intervals around $\mathbf{x}$, $\mathbf{v}$, and $m$, respectively, at time $t$,
\begin{eqnarray}
 \d N=f(\mathbf{x},\mathbf{v},m,t)\d^3x\d^3v\d m. \label{def_f}
\end{eqnarray}
Moreover, $f$ determines the mass density,
\begin{eqnarray}
 \rho(\mathbf{x},t)=\left\langle\int\limits_{\mathbb{R}^4}mf(\mathbf{x},\mathbf{v},m,t)\d^3v\d m\right\rangle_\mathrm{t}, \label{def_rho}
\end{eqnarray}
while the mass-weighted average of a quantity $A(\mathbf{x},\mathbf{v},m,t)$ is defined through the equation,
\begin{equation}
 \langle A\rangle(\mathbf{x},t)=\frac{1}{\rho(\mathbf{x},t)}\left\langle\int\limits_{\mathbb{R}^4}m(Af)(\mathbf{x},\mathbf{v},m,t)\d^3v\d m\right\rangle_\mathrm{t}. \label{def_av}
\end{equation}
In Eqs.~(\ref{def_rho}) and (\ref{def_av}) $\langle\cdot\rangle_\mathrm{t}$ denotes the time average,
\begin{eqnarray}
 \langle A\rangle_\mathrm{t}=\lim_{T\rightarrow\infty}\frac{1}{T}\int\limits_t^{t+T}A(t')\d t'. \label{time_av}
\end{eqnarray}

Using these definitions, we can calculate the total transported mass per unit soil area ($M$), the total mass flux ($Q$), and the average particle velocity ($V$) from the expressions,
\begin{eqnarray}
 M&=&\int\limits_{h_o}^\infty\rho\d z, \label{M} \\
 Q&=&\int\limits_{h_o}^\infty\rho\langle v_x\rangle\d z=M\overline{\langle v_x\rangle}, \label{defQ} \\
 V&=&\frac{Q}{M}=\overline{\langle v_x\rangle}, \label{defV}
\end{eqnarray}
respectively, where the overbar denotes the mass-weighted height average,
\begin{eqnarray}
 \overline A&=&\frac{\int\limits_{h_o}^\infty\rho A\d z}{\int\limits_{h_o}^\infty\rho\d z}=\frac{1}{M}\int\limits_{h_o}^\infty\rho A\d z. \label{heightav}
\end{eqnarray}

\subsection{\label{mass_and_momentum}Local mass and momentum conservation equations}
In this Section, the local average mass and momentum conservation equations for our particle system are presented using the notations and definitions introduced in the last Section. The derivation of these conservation equations can be found in Babic \cite{Babic97}. For our system ($\partial/\partial t=\partial/\partial y=0$), these equations are,
\begin{eqnarray}
 &&\frac{\partial\rho\langle v_x\rangle}{\partial x}+\frac{\partial\rho\langle v_z\rangle}{\partial z}=0, \label{massb} \\
 &&\frac{\partial}{\partial x}(\rho\langle v_x^2\rangle+P_{xx})=\rho\langle a^{\mathrm{ex}}_x\rangle-\frac{\partial}{\partial z}(\rho\langle v_xv_z\rangle+P_{xz}), \label{mombx} \\
 &&\frac{\partial}{\partial x}(\rho\langle v_xv_z\rangle+P_{zx})=\rho\langle a^{\mathrm{ex}}_z\rangle-\frac{\partial}{\partial z}(\rho\langle v_z^2\rangle+P_{zz}), \label{mombz}
\end{eqnarray}
where $P_{ij}$ is given by \cite{Babic97},
\begin{eqnarray}
 && P_{ij}=\frac{1}{2}\left\langle\sum_{mn}F_i^{mn}x_j^{nm}\int\limits_0^1\delta(\mathbf{x}-\mathbf{x}^n-s\mathbf{x}^{nm})\d s\right\rangle_\mathrm{t}, \nonumber \\
 && \label{Pij}
\end{eqnarray}
with $\mathbf{x}^{mn}=\mathbf{x}^n-\mathbf{x}^m$. $P_{ij}$ is the contact force contribution to the particle stress tensor since its gradient compensates the contact force density \cite{Babic97},
\begin{eqnarray}
 \frac{\partial P_{ij}}{\partial x_j}=-\left\langle\sum_{mn}F_i^{mn}\delta(\mathbf{x}-\mathbf{x}^m)\right\rangle_\mathrm{t}.
\end{eqnarray}
It describes the momentum flux due to collisions between particles. In fact, even though the total momentum is conserved in collisions, the finite size of the particles and thus $\mathbf{x}^{mn}\ne0$ lead to a shift of the location of this momentum. We note that this shift of the momentum location in collisions has been neglected in our model derivation in Ref.~\cite{Paehtzetal13} (dilute approximation). As a consequence, Eq.~(\ref{mombx}) is a generalization of Eq.~(2) of Ref.~\cite{Paehtzetal13}, such that these two equations are equal if the contributions from $P_{ij}$ in Eq.~(\ref{mombx}) are neglected. The distribution $\int_0^1\delta(\mathbf{x}-\mathbf{x}^n-s\mathbf{x}^{nm})\d s$ appearing in Eq.~(\ref{Pij}) is the mathematical expression for a ''delta line`` between $\mathbf{x}^m$ and $\mathbf{x}^n$. Integrating this distribution over an arbitrary domain yields the fraction of the line contained in this domain. The inhomogeneities introduced by this and the other delta distributions indirectly appearing in quantities of the type $\rho\langle\cdot\rangle$ are smoothed out by the time averaging procedure $\langle\cdot\rangle_\mathrm{t}$, which is also incorporated in the definition of $\langle\cdot\rangle$.

\subsection{\label{sec:saturation}Differential equations of flux saturation}
The results of the last Section can be now used in order to derive the saturation equations for the average transported mass per unit soil area ($M$) and the average particle velocity ($V$), used to define the sediment flux, $Q=MV$. To do so, we first integrate Eqs.~(\ref{massb})-(\ref{mombz}) over the height. This calculation is the subject of Section \ref{height_integration}. Thereafter, in Section \ref{momentum_balance}, we combine the resulting horizontal and vertical momentum balances by means of a Coulomb friction law and rewrite each term of the horizontal momentum balance equation in terms of $M$ and $V$. We then present the mass and horizontal momentum balance equations in their final form in Section \ref{final_form}.

\subsubsection{Height-integrated conservation equations}\label{height_integration}
Since our description relates to the saturation of the mass flux $Q$ due to changes in the downstream direction ($x$), we integrate Eqs.~(\ref{massb})-(\ref{mombz}) over height ($\int_{h_o}^\infty\cdot\d z$). By using Eqs.~(\ref{M})-(\ref{heightav}) and by further taking into account $\partial h_o/\partial x\approx0$ and $\rho(\infty)=0$, this height-integration yields,
\begin{eqnarray}
 && \frac{\d}{\d x}(MV)=(\rho\langle v_z\rangle)(h_o), \label{massb2} \\
 && \frac{\d}{\d x}(M\overline{\langle v_x^2\rangle+P_{xx}/\rho})=M\overline{\langle a_x^\mathrm{ex}\rangle}+(\rho\langle v_xv_z\rangle+P_{xz})(h_o), \nonumber \\
 && \label{mombalance3x_new} \\
 && \frac{\d}{\d x}(M\overline{\langle v_xv_z\rangle+P_{zx}/\rho})=M\overline{\langle a_z^\mathrm{ex}\rangle}+(\rho\langle v_z^2\rangle+P_{zz})(h_o). \nonumber \\
 && \label{mombalance3z}
\end{eqnarray}
We note that Eq.~(\ref{mombalance3x_new}) corresponds to Eq.~(2) of Ref.~\cite{Paehtzetal13} if the contributions from $P_{ij}$ in Eq.~(\ref{mombalance3x_new}) are neglected (dilute approximation).

{\em{Coulomb friction law}} --- The terms $(\rho\langle v_xv_z\rangle+P_{xz})(h_o)$ and $(\rho\langle v_z^2\rangle+P_{zz})(h_o)$ are the vertical fluxes of horizontal and vertical momentum component per unit volume at the location of the sediment bed, respectively, whereby the velocity terms are the contributions due to particle motion, and $P_{xz}(h_o)$ and $P_{zz}(h_o)$ are the contributions due to collisional momentum transfer. In other words, these two terms describe the total amounts of horizontal and vertical momentum, respectively, per unit soil area that enter the transport layer per unit time from the sediment bed. These momentum changes per unit area and time of the transport layer can be seen as being caused by an effective force per unit area ($\mathbf{f}^\mathrm{bed}$) which the sediment bed applies on the transport layer,
\begin{eqnarray}
 f_x^\mathrm{bed}&=&(\rho\langle v_xv_z\rangle+P_{xz})(h_o), \\
 f_z^\mathrm{bed}&=&(\rho\langle v_z^2\rangle+P_{zz})(h_o).
\end{eqnarray}
Bagnold \cite{Bagnold56,Bagnold73} was the first who proposed that these force components are related to each other through a Coulomb friction law, independent of whether the transport regime is subaqueous or aeolian. That is,
\begin{eqnarray}
 f_x^\mathrm{bed}=-\mu f_z^\mathrm{bed}, \label{Coulombassumption}
\end{eqnarray}
where $\mu$ is the Coulomb friction coefficient. Models for saturated sediment transport using this Coulomb friction law have been successfully validated through comparison with experiments, thus giving support to the Coulomb friction law adapted to sediment transport (e.g. \cite{Bagnold56,AshidaMichiue72,Bagnold73,Paehtzetal12a}). Additional support comes from numerical simulations of saturated ($\partial/\partial x=0$) granular Couette flows under gravity. Zhang and Campbell \cite{ZhangCampbell92} found for such flows that the interface between the particle bed and the transport layer is characterized by a constant ratio between the $xz$ and $zz$ components of the particle stress tensor ($\mathbf{T}$), $T_{xz}=-\mu T_{zz}$. Since both Couette flow and sediment transport along the surface are granular shear flows, it seems reasonable that also the interface between the sediment bed and the transport layer for saturated sediment transport along the surface is characterized by such a law. Indeed, $f_x^\mathrm{bed}$ and $f_z^\mathrm{bed}$ become equal to $T_{xz}$ and $T_{zz}$, respectively, if $\langle v_z\rangle=0$ \cite{Babic97}, which is fulfilled for saturated sediment transport since $\partial/\partial x=0$ implies $\d(\rho\langle v_z\rangle)/\d z=0$ (cf. Eq.~(\ref{massb})), which in turn implies $\langle v_z\rangle=0$ due to $\rho\langle v_z\rangle$ vanishing sufficiently deep within the sediment bed. Finally, it seems reasonable that the Coulomb friction law should be also approximately valid in situations weakly out-of-equilibrium \cite{Sauermannetal01}, provided the sediment flux is close to its saturated value. Assuming the validity of Eq.~(\ref{Coulombassumption}), we can combine Eqs.~(\ref{mombalance3x_new}) and (\ref{mombalance3z}) to,
\begin{eqnarray}
 \frac{\d}{\d x}(c_vMV^2)=M\overline{\langle a_x^\mathrm{ex}\rangle}+\mu M\overline{\langle a_z^\mathrm{ex}\rangle}, \label{mombalance3x}
\end{eqnarray}
where $c_v$ is a correlation factor given by,
\begin{eqnarray}
 c_v=\frac{1}{V^2}\overline{\langle v_x^2\rangle+P_{xx}/\rho+\mu(\langle v_xv_z\rangle+P_{zx}/\rho)}. \label{cvdefprecise}
\end{eqnarray}
{\em{The correlation factor}} --- Since we are only interested in situations close to equilibrium, and since at equilibrium $\langle v_z\rangle=0$ (see discussion in the previous paragraph), it follows $\langle v_x^2\rangle\gg\mu|\langle v_xv_z\rangle|$ ($\mu$ is of order unity). Moreover, for sufficiently dilute granular flows, the momentum transfer in collisions is small and thus $\langle v_x^2\rangle\gg |P_{ij}|/\rho$. While sediment transport in the aeolian regime is certainly dilute enough to ensure this condition for most of the transport layer, sediment transport in the subaqueous regime might not fulfill it because a large part of the transport occurs in rather dense regions of the transport layer \cite{Duranetal12}. However, using the code of Dur\'an et al. \cite{Duranetal12}, we confirmed that $\langle v_x^2\rangle\gg |P_{ij}|/\rho$ also for subaqueous transport. Hence, $c_v$ can be approximated as,
\begin{eqnarray}
 c_v\approx\frac{\overline{\langle v_x^2\rangle}}{V^2}. \label{cvdef}
\end{eqnarray}
We confirmed, using the code of Dur\'an et al. \cite{Duranetal12}, that for transport in equilibrium ($\partial/\partial x=0$) $c_v$ is nearly constant with the fluid shear velocity, $u_{\ast}$, in both sediment transport regimes. Hence, it seems reasonable that changes of $c_v$ with $x$ during the saturation process of the sediment flux close to equilibrium can be regarded as negligible compared to the corresponding changes of $M$ or $V$ with $x$. In this manner, we can consider the value of $c_v$ associated with sediment transport in equilibrium, independent of the downstream position and of the fluid shear velocity. This leads to the following approximation for $c_v$,
\begin{equation}
c_v\approx\frac{{\overline{\langle v_x^2\rangle}_{\!s}}}{V_s^2}. \label{cvdef2}
\end{equation}
where ${\overline{\langle v_x^2\rangle}_{\!s}}$ is the equilibrium value of ${\overline{\langle v_x^2\rangle}}$. This equilibrium value of $c_v$ can be determined from experiments as we will discuss in Section \ref{cv_aeolian} and \ref{cv_subaqueous}.

\subsubsection{Momentum balance equation in terms of $M$ and $V$}\label{momentum_balance}

Now we express both terms on the right-hand-side of the momentum conservation equation, i.e. Eq.~(\ref{mombalance3x}), as functions of $M$ and $V$ in order to obtain a differential equation describing the saturation of $M$ and $V$. 

The first term on the right-hand-side of Eq.~(\ref{mombalance3x}) can be written as \cite{Sauermannetal01,Paehtzetal12a},
\begin{eqnarray}
 M\overline{\langle a_x^\mathrm{ex}\rangle}=\frac{3M}{4c_asd}\cdot C_d(V_r)\cdot V_r^2, \label{drag}
\end{eqnarray}
where $s=\rho_p/\rho_f$ is the ratio between sediment and fluid density; $V_r$ is defined as,
\begin{equation}
V_r=U-V, \label{Vr}
\end{equation}
which is the difference between the average fluid velocity ($U=\overline{u}$) and the average horizontal particle velocity ($V$), where $u(z)$ is the fluid velocity profile; $C_d$ is the drag coefficient, which is a function of $V_r$, and $c_a$ accounts for the added mass force through,
\begin{equation}
c_a=1+\frac{1}{2s}. \label{ca}
\end{equation}
The added mass force arises when the particle is accelerated relative to the surrounding fluid, because the fluid layer immediately surrounding the particle will also be accelerated. As denoted by Eq. (\ref{ca}), this ``added mass'' of the fluid layer amounts to approximately one half the weight of the fluid displaced by the particle \cite{VanRijn93}. While the added mass correction is significant for transport in a dense medium such as water \cite{NinoGarcia98a}, it is negligibly small for sediment transport in the aeolian regime since $c_a\approx1$ for large $s$. Thus, this correction is usually disregarded in studies of aeolian sediment transport (e.g. Refs.~\cite{Sauermannetal01,Paehtzetal12a}). We note that Eq.~(\ref{drag}) is not valid for dense transport regimes like sheet flow, in which the drag coefficient displays a strong dependence on the concentration profile of transported particles \cite{Duetal06}. In this manner, Eq.~(\ref{drag}) can be used in the present study because our analytical treatment considers the two main modes of transport, namely saltation and creep.

The second term on the right-hand-side of Eq.~(\ref{mombalance3x}), $\mu M\overline{\langle a_z^\mathrm{ex}\rangle}$, can be taken as approximately equal to the buoyancy-reduced gravity force \cite{Bagnold73,Paehtzetal12a} corrected by the added mass force. It can be written as,
\begin{eqnarray}
 \mu M\overline{\langle a_z^\mathrm{ex}\rangle}=-\frac{\mu}{c_a}\tilde gM, \label{friction}
\end{eqnarray}
where $\tilde g=(s-1)g/s$ is the buoyancy-reduced value of the gravity constant, $g$.

\subsubsection{The conservation equations in their final form}\label{final_form}

By substituting Eqs.~(\ref{drag}) and (\ref{friction}) into Eq.~(\ref{mombalance3x}) using $\d c_v/\d x\approx0$ (cf. Eq.~(\ref{cvdef2})), we obtain the momentum conservation equation in terms of $M$ and $V$, whereas Eq.~(\ref{massb2}) gives the mass balance. Therefore, the mass and momentum conservation equations in their final form read,
\begin{eqnarray}
 \frac{\d(MV)}{\d x}&=&(\rho\langle v_z\rangle)(h_o), \label{massconserv} \\
 c_{v}\frac{\d(MV^2)}{\d x}&=&\frac{3M}{4c_asd}\cdot C_d(V_r)\cdot V_r^2-\frac{\mu}{c_a}\tilde gM. \label{momconserv}
\end{eqnarray}
We note that Eq.~(\ref{momconserv}) is identical to Eq.~(4) of Ref.~\cite{Paehtzetal13} if the definition of $c_a$ (Eq.~(\ref{ca})) is inserted. We further note that Eq.~(\ref{momconserv}) can be used to obtain the saturated value $V_{rs}$ of the velocity difference $V_r$. By using $\d/\d x=0$ (saturated sediment transport), we obtain,
\begin{eqnarray}
 \frac{3}{4sd}\cdot C_d(V_{rs})\cdot V_{rs}^2=\mu\tilde g, \label{vrs}
\end{eqnarray}
which can be numerically solved for $V_{rs}$.

\section{Obtaining the flux saturation length of sediment transport}\label{Obtaining_Ls}

In this Section, we use the results presented in last Section in order to derive a closed expression for the saturation length as a function of the attributes of sediment and flow, both for aeolian and subaqueous regimes. The derivation of the saturation length equation is the subject of Section \ref{derivation_of_Ls}. In Section \ref{equations_for_Ls} we present and discuss the resulting equation for the saturation length. In Sections \ref{Ls_aeolian} and \ref{Ls_subaqueous} we show how the saturation length equation can be applied to compute $L_s$ in the aeolian and subaqueous regimes, respectively.

\subsection{Derivation}\label{derivation_of_Ls}

Close to equilibrium, $M$ and $V$ saturate simultaneously following a certain function $M(V)$, where $M_s=M(V_s)$. This function is linked to the characteristics of the erosion and deposition of bed material and thus to the unknown shape of $(\rho\langle v_z\rangle)(h_o)$ as a function of $M$ and $V$ in Eq.~(\ref{massconserv}). Moreover, also the mean fluid velocity $U$ will saturate following a certain function $U(V)$ close to the saturated regime, since $U$ is influenced by the feedback of the sediment transport on the fluid flow. Therefore, Eq.~(\ref{Vr}) becomes,
\begin{equation}
V_r(V)=U(V)-V. \label{VrV}
\end{equation}
By taking into account that both $M$ and $U$ are functions of $V$, and by using Eq.~(\ref{vrs}), we can rewrite the momentum balance Eq.~(\ref{momconserv}) in such a way to obtain the following expression for ${\d V}/{\d x}$, 
\begin{equation}
 \frac{\d V}{\d x}=\Omega(V)=A(V)\cdot B[V_r(V)], \label{Omegadef}
\end{equation}
where the functions $A(V)$ and $B(V)$ are given by the equations,
\begin{eqnarray}
&&A(V)=\frac{3M\!(V)}{4sdc_ac_v\left(2V\cdot M\!(V)+V^2\frac{\d M\!(V)}{\d V}\right)}; \nonumber \\
&&\label{A_V} \\
&&B(V_r)=C_d(V_r)\cdot V_r^2-C_d(V_{rs})\cdot V_{rs}^2. \label{B_V}
\end{eqnarray}
Furthermore, since $Q(V)=M(V)V$, we obtain,
\begin{eqnarray}
 \frac{\d V}{\d Q}(V)=\left(M(V)+V\frac{\d M(V)}{\d V}\right)^{-1}. \label{dV_dQ}
\end{eqnarray}
In this manner, using Eq.~(\ref{Omegadef}), $\Gamma(V)$ can be written as,
\begin{eqnarray}
 \Gamma(V)=\frac{\d Q}{\d x}(V)=\left(M(V)+V\frac{\d M(V)}{\d V}\right)\Omega(V). \label{Gamma_V}
\end{eqnarray}
Using Eqs.~(\ref{dV_dQ}) and (\ref{Gamma_V}), we can write Eq.~(\ref{Lsdef2}) for the saturation length as,
\begin{eqnarray}
 L_s=-\left(\frac{\d\Gamma}{\d V}\frac{\d V}{\d Q}\right)_{V=V_s}^{-1}=-\left(\frac{\d\Omega}{\d V}\right)_{V=V_s}^{-1}, \label{LV}
\end{eqnarray}
where we further used that $\Omega(V_s)=0$. 

Calculating $L_s$ through Eq.~(\ref{LV}) requires obtaining an expression for $\d\Omega/\d V$, where $\Omega(V)$ is defined in Eq.~(\ref{Omegadef}). However, $\Omega(V)$ incorporates, through the function $B(V)$ defined in Eq.~(\ref{B_V}), a dependence on the equilibrium value of the relative velocity $V_r$, i.e. $V_{rs}$. In order to obtain an expression for $V_{rs}$, we solve Eq.~(\ref{vrs}) for $V_{rs}$ using the drag law of Julien \cite{Julien95} for natural sediment, which writes,
\begin{eqnarray}
 C_d(v_r)=\frac{24\nu}{v_rd}+1.5, \label{dragjulien}
\end{eqnarray}
whereas we find that the specific choice of the drag law has only a small effect on the value of $L_s$ obtained from our calculations. By substituting the expression for $C_d(V_{rs})$, obtained with Eq.~(\ref{dragjulien}), into Eq.~(\ref{vrs}), and solving this equation for $V_{rs}$, yields,
\begin{eqnarray}
 V_{rs}=\frac{\sqrt{8\mu s\tilde gd^3+(24\nu)^2}-24\nu}{3d}. \label{vrs2}
\end{eqnarray}
This equation is, then, used to compute $B(V_r)$ through Eq.~(\ref{B_V}), whereupon $\Omega(V)$ can be obtained using Eqs.~(\ref{VrV}) and (\ref{Omegadef}). The resulting expression for the saturation length, computed with Eq.~(\ref{LV}), reads,
\begin{equation}
 L_s=\frac{(2+c_M)c_ac_vV_{rs}V_sF}{\mu\tilde g}\cdot \left(1-\frac{\d U}{\d V}(V_s)\right)^{-1}, \label{LVfinal}
\end{equation}
where the quantity,
\begin{eqnarray}
 c_M=\frac{V_s}{M_s}\frac{\d M}{\d V}(V_s), \label{cMdef}
\end{eqnarray}
describes the relative change of $M$ with $V$ close to the saturated regime, while $F$ is given by,
\begin{eqnarray}
 F=C_d(V_{rs})\cdot V_{rs}\cdot\left(\frac{\d(C_dV_r^2)}{\d V_r}\right)^{-1}_{V_r=V_{rs}}= \nonumber \\ \frac{24V_{rs}\nu/d+1.5V_{rs}^2}{24V_{rs}\nu/d+3V_{rs}^2}=\frac{V_{rs}+16\nu/d}{2V_{rs}+16\nu/d}, \label{Cd_Fx}
\end{eqnarray}
and thus $F$ encodes information about the drag law.

In order to obtain our final expression for $L_s$, we need to express $\frac{\d U}{\d V}(V_s)$. We note that, for the saturated state, the mean flow velocity $U$ is dominantly a function of the shear velocity $u_{\ast}$ and the shear velocity at the bed \cite{Paehtzetal12a,Duranetal12}, that is,
\begin{equation}
 u_b=\sqrt{\tau_f(h_o)/\rho_f}. \label{ub}
\end{equation}
The shear velocity at the bed, $u_b$, is reduced due to the feedback of the sediment transport on the fluid flow, where $\tau_f(z)$ is the fluid shear stress profile. We can express $u_b$ using the inner turbulent boundary layer approximation of the Navier-Stokes equations. These equations approximate the Navier-Stokes equations for heights $z$ much smaller than the height $\delta_b$ of the boundary layer, which is the region in which we are interested. George \cite{George09} derived the inner boundary layer approximation of the Navier-Stokes equations in the absence of an external body force. In the presence of an external body force, these equations must be slightly modified by adding the body force term in the momentum equations. The horizontal momentum equation thus writes \cite{George09},
\begin{eqnarray}
 \frac{\d\tau_f}{\d z}=-F_{x\mathrm{body}}, \label{boundarylayer}
\end{eqnarray}
where $F_{x\mathrm{body}}(z)$ is the horizontal body force per unit volume acting on the flow at each height $z$. $F_{x\mathrm{body}}$ is the drag force per unit volume which the particles apply on the fluid. In other words, $F_{x\mathrm{body}}$ is the reaction force per unit volume of the horizontal force per unit volume which the fluid applies on the particles. That is,
\begin{eqnarray}
 F_{x\mathrm{body}}=-\rho\langle a_x^\mathrm{ex}\rangle. \label{Fxbody}
\end{eqnarray}
We then substitute Eq.~(\ref{Fxbody}) into Eq.~(\ref{boundarylayer}) and integrate this equation from $z=h_o$ to $z=z_\mathrm{cut}$, where $z_\mathrm{cut}\ll\delta_b$ is a height which incorporates the entire transport layer, thereby using $\tau_f(z_\mathrm{cut})=\tau=\rho_fu_{\ast}^2$, and $\int_{h_o}^{z_\mathrm{cut}}\rho\langle a_x^\mathrm{ex}\rangle=\int_{h_o}^\infty\rho\langle a_x^\mathrm{ex}\rangle=M\overline{\langle a_x^\mathrm{ex}\rangle}$. This leads to,
\begin{equation}
 \tau_f=\tau-M\overline{\langle a_x^\mathrm{ex}\rangle}. \label{tautauf}
\end{equation}
By substituting this equation into Eq.~(\ref{ub}) and using Eq.~(\ref{drag}), we obtain the following equation for $u_b$,
\begin{equation} 
u_b=u_{\ast}\sqrt{1-\frac{M\overline{\langle a_x^\mathrm{ex}\rangle}}{\rho_fu_{\ast}^2}}=u_{\ast}\sqrt{1-\frac{\frac{3M}{4sdc_a}C_d(V_r)V_r^2}{\rho_fu_{\ast}^2}}. \label{ubV}
\end{equation}
Since $u_{\ast}$ does not depend on $V$, we can now express $\frac{\d U}{\d V}(V_s)$ as,
\begin{eqnarray}
 \frac{\d U}{\d V}(V_s)=c_U\frac{V_s+V_{rs}}{u_{bs}}\frac{\d u_b}{\d V}(V_s), \label{UVs}
\end{eqnarray}
where $u_{bs}$ is the value of $u_b$ in equilibrium, and the quantity $c_U$ is given by the equation,
\begin{eqnarray}
 c_U=\frac{u_{bs}}{U_s}\frac{\d U}{\d u_b}(u_{bs})=\frac{u_{bs}}{V_s+V_{rs}}\frac{\d U}{\d u_b}(u_{bs}). \label{cUdef}
\end{eqnarray}
where we used, $V_{rs}=U_s-V_s$. We note that $c_U$ describes the relative change of $U$ with $u_b$ close to the saturated regime. Moreover, the derivative $\frac{\d u_b}{\d V}(V_s)$ can be calculated using Eq.~(\ref{ubV}) with, $M=M(V)$, $V_r=V_r(V)$, and,
\begin{eqnarray}
 \frac{\frac{3M_s}{4sdc_a}\cdot C_d(V_{rs})\cdot V_{rs}^2}{\rho_fu_{\ast}^2}=1-\frac{u_{bs}^2}{u_{\ast}^2},
\end{eqnarray}
which follows from $u_b(V_s)=u_{bs}$. We thus obtain,
\begin{eqnarray}
 \frac{\d u_b}{\d V}(V_s)=\frac{u_{\ast}^2-u_{bs}^2}{2u_{bs}}\cdot\left[\frac{1-\frac{\d U}{\d V}(V_s)}{FV_{rs}}-\frac{c_M}{V_s}\right].
\end{eqnarray}
This expression is substituted into Eq.~(\ref{UVs}), whereas the resulting equation is then solved for $\left(1-\frac{\d U}{\d V}(V_s)\right)^{-1}$ --- this is the term involving $\d U / \d V$ which we need to compute $L_s$ in Eq.~(\ref{LVfinal}). In this manner, we finally obtain a closed expression for the saturation length, which we present and discuss in the next subsection.

\subsection{The saturation length equation}\label{equations_for_Ls}
The equation for the saturation length, $L_s$, which is identical to Eq.~(5) of Ref.~\cite{Paehtzetal13} if the definition of $c_a$ (Eq.~(\ref{ca})) is inserted, reads,
\begin{equation} 
L_s=\frac{(2+c_M)c_ac_vV_{rs}V_sFK}{\mu\tilde g}, \label{LVfinal2} 
\end{equation}
where $V_{rs}$ and $F$ are calculated using Eqs.~(\ref{vrs}) and (\ref{Cd_Fx}), respectively. In addition, the last factor $K$ on the right-hand-side of Eq.~(\ref{LVfinal2}) is given by the equation,
\begin{widetext}
\begin{equation}
 K=\left(1-\dfrac{\d U}{\d V}(V_s)\right)^{-1}=\dfrac{1+\left[\dfrac{c_U\cdot(V_s+V_{rs})}{2FV_{rs}}\right]\cdot\left(\dfrac{u_{\ast}^2}{u_{bs}^2}-1\right)}{1+\left[\dfrac{c_Uc_M\cdot(V_s+V_{rs})}{2V_s}\right]\cdot\left(\dfrac{u_{\ast}^2}{u_{bs}^2}-1\right)}\approxeq\dfrac{1+\left[\dfrac{c_U\cdot(V_s+V_{rs})}{2FV_{rs}}\right]\cdot\left(\dfrac{u_{\ast}^2}{u_{{\mathrm{t}}}^2}-1\right)}{1+\dfrac{c_Uc_M\cdot(V_s+V_{rs})}{2V_s}\cdot\left(\dfrac{u_{\ast}^2}{u_{{\mathrm{t}}}^2}-1\right)}, \label{Lsfluid}
\end{equation}
\end{widetext}
and $K$ thus encodes information about the saturation of the transport-flow feedback. In fact, if the saturation of the transport-flow feedback is neglected ($U=U_s$), it follows $c_U=0$ and thus $K=1$. We note that Eq.~(\ref{Lsfluid}) is identical to Eq.~(9) of Ref.~\cite{Paehtzetal13}, which we obtained for aeolian transport, for $c_M\approx c_U\approx1$ (see Section \ref{Ls_aeolian}). Moreover, for transport in the subaqueous regime, $c_U\approx0$ as shown in Section \ref{Ls_subaqueous}. Therefore, in this regime, Eq.~(\ref{Lsfluid}) gives $K\approx1$, which is the result we obtained for subaqueous transport in Ref.~\cite{Paehtzetal13}. 
In fact, using the corresponding values for $c_M$ and $c_U$ and inserting Eq.~(\ref{ca}), Eq.~(\ref{LVfinal2}) becomes equal to,
\begin{eqnarray}
 L_s^{\mathrm{subaq}}=\frac{(2s+1)c_v{V_s}V_{rs}F}{\mu(s-1)g}, \label{LVfinal2subq}
\end{eqnarray}
for subaqueous transport and,
\begin{eqnarray}
L_s^{\mathrm{aeolian}}=\frac{3c_vV_sV_{rs}FK}{\mu g}, \label{LVfinal2aeol}
\end{eqnarray}
for aeolian transport, where we further used $(s+0.5)/(s-1)\approx1$ for aeolian transport. Eqs.~(\ref{LVfinal2subq}) and (\ref{LVfinal2aeol}) are identical to Eqs.~(8) and (10) of Ref.~\cite{Paehtzetal13}, respectively.

In Eq.~(\ref{Lsfluid}), we assumed that the saturated shear velocity at the bed ($u_{bs}$) and the bed shear stress in equilibrium ($\tau_{fs}(h_o)$) approximately equal $u_{{\mathrm{t}}}$ and $\tau_{\mathrm{t}}$, respectively, i.e. the minimal shear velocity and the minimal shear stress at which sediment transport can be sustained,
\begin{eqnarray}
\tau_{fs}(h_o)=\tau_{\mathrm{t}}, \label{tauftaut} \\
u_{bs}=u_{{\mathrm{t}}}. \label{ubut}
\end{eqnarray}
In the following, we present arguments which justify this assumption. 

For aeolian sediment transport, Eqs.~(\ref{tauftaut}) and (\ref{ubut}) are known as ``Owen's hypothesis''. These equations are known to be approximately valid when $u_{\ast}$ is close to the threshold (e.g. Figure 2.10 in Ref. \cite{Koketal12}). However, as $u_{\ast}$ increases, $u_{bs}$ actually decreases away from $u_{{\mathrm{t}}}$ \cite{Paehtzetal12a,Koketal12}. Nonetheless, the approximation which we use in Eq.~(\ref{Lsfluid}) is reasonable even for large shear velocities, since, when $u_{\ast}$ is significantly larger than $u_{{\mathrm{t}}}$ (which means $u_{\ast}>2u_{{\mathrm{t}}}$ for Earth conditions with $c_M=c_U=1$), we have that,
\begin{equation}
 K\approxeq\frac{V_s}{c_MF V_{rs}}, \label{Kasym}
\end{equation}
which is nearly independent of $u_{bs}$. Using this approximation with $c_M\approx1$, Eq.~(\ref{LVfinal2aeol}) becomes,
\begin{eqnarray}
L_s^{\mathrm{aeolian}}=\frac{3c_vV_s^2}{\mu g}, \label{LVfinal2aeolapp}
\end{eqnarray}
which is identical to Eq.~(39) of the supplementary material of Ref.~\cite{Paehtzetal13}.

For subaqueous sediment transport, Eqs.~(\ref{tauftaut}) and (\ref{ubut}) are known as ``Bagnold's hypothesis''. This hypothesis is widely used in the literature (e.g. \cite{Bagnold56,AshidaMichiue72,Bagnold73,NinoGarcia94}), although some studies have questioned it (e.g. \cite{NinoGarcia98a,Seminaraetal02}). However, there is evidence from recent studies that this hypothesis is approximately fulfilled. In order to review this evidence, we use Eqs.~(\ref{drag}), (\ref{vrs}), and (\ref{tautauf}) to express $M_s$ as,
\begin{eqnarray}
 M_s=\frac{c_a}{\mu\tilde g}\cdot \left[\tau-\tau_{fs}(h_o)\right]. \label{Ms}
\end{eqnarray}
To our knowledge, the only study in which $M_s$ has been measured as a function of $\tau$ is the recent study of Lajeunesse et al. \cite{Lajeunesseetal10}, who obtained, using video-imaging techniques, that,
\begin{eqnarray}
 M_s=\frac{1}{0.415\tilde g}\cdot \left[\tau-\tau_{\mathrm{t}}\right]. \label{Msshape}
\end{eqnarray}
Therefore, if we assume $\tau_{fs}(h_o)=\tau_{\mathrm{t}}$ as in Eq.~(\ref{tauftaut}), then, by comparing Eqs.~(\ref{Ms}) and (\ref{Msshape}) with $c_a=1.19$ valid for subaqueous sediment transport (cf. Eq. (\ref{ca}) with $s=2.65$), we obtain $\mu/c_a=0.415$ and thus $\mu\approxeq0.493$. Indeed, values within the range between $\mu/c_a=0.3$ and $\mu/c_a=0.5$ --- and thus consistent with the value of $\mu$ estimated above --- have been reported from measurements of particle trajectories in the subaqueous sediment transport \cite{Francis73,AbbottFrancis77,Ninoetal94}. Further evidence that Bagnold's hypothesis is approximately correct was provided by the recent numerical study of Dur\'an et al. \cite{Duranetal12}. These authors simulated the dynamics of both the sediment bed and of transported particles at the single particle scale. Dur\'an et al. \cite{Duranetal12} found that $M_s\tilde g\propto(\tau-\tau_{\mathrm{t}})$, which is similar to Eq.~(\ref{Msshape}) and can satisfactorily explain all simulated data with a single proportionality constant. Moreover, the authors also found that $\tau_{fs}$ reduces to $\tau_{\mathrm{t}}$ at a height $z$ very close to the top of the bed, $z\approxeq h_o$. Given these separate lines of evidence, we believe that Bagnold's hypothesis is a reasonable approximation. Moreover, we emphasize that our analysis for subaqueous sediment transport is not affected by this approximation, since we estimate in Section \ref{cu_subaqueous} that $c_U\approx0$ and thus $K\approx1$ regardless of the value of $u_{bs}$.

In summary, the saturation length of sediment flux, $L_s$, can be calculated using Eq.~(\ref{LVfinal2}), where $V_{rs}$ and $F$ are given by Eqs.~(\ref{vrs2}) and (\ref{Cd_Fx}), respectively, while Eq.~($\ref{Lsfluid}$) is used to compute the term, $\left(1-\frac{\d U}{\d V}(V_s)\right)^{-1}$, which appears on the right-hand-side of Eq.~(\ref{LVfinal2}). These equations include certain quantities which depend on the characteristics of the sediment transport and thus on the transport regime. These quantities are $c_v$, $c_M$, $c_U$, $\mu$, the saturated particle velocity $V_s$, and the threshold shear velocity, $u_{{\mathrm{t}}}$. We estimate these quantities for the aeolian regime of sediment transport in Section \ref{Ls_aeolian} and for the subaqueous regime in Section \ref{Ls_subaqueous}.

\subsection{The saturation length of aeolian sediment transport}\label{Ls_aeolian}
In this section, we estimate the parameters $c_v$, $c_U$, $c_M$, $\mu$, and express the saturated particle velocity $V_s$ and the threshold shear velocity $u_{{\mathrm{t}}}$ for aeolian sediment transport. Note that we estimate these parameters only roughly, which is sufficient in the light of the large scatter (factor $2-4$) of the experimental data \cite{Andreottietal10,Fourriereetal10}.

\subsubsection{The parameter $c_v$}\label{cv_aeolian}
In this section, we reiterate some of the results we obtained in Section~A1 of the supplementary material of Ref.~\cite{Paehtzetal13}. The parameter $c_v$ (Eq.~(\ref{cvdef})) occurs as a prefactor in Eq.~(\ref{LVfinal2}), and thus determines the magnitude of $L_s$. Since,
\begin{eqnarray}
 \overline{\langle(v_x-\overline{\langle v_x\rangle})^2\rangle}=\overline{\langle v_x^2\rangle}-\overline{\langle v_x\rangle}^2>0,
\end{eqnarray}
we conclude that $c_v$ must be larger than unity, that is,
\begin{eqnarray}
 c_v=\frac{\overline{\langle v_x^2\rangle}}{\overline{\langle v_x\rangle}^2}>1.
\end{eqnarray}
However, experiments on aeolian sediment transport \cite{Creysselsetal09} show that the change of $\langle v_x\rangle(z)$ with $z$ is small close to $h_o$, where most of the transport takes place. Consequently, the value of $c_v$ must be close to unity. 

We estimate $c_v$ from experiments on aeolian sediment transport \cite{Greeleyetal96,Creysselsetal09}. Creyssels et al. \cite{Creysselsetal09} measured an exponentially decaying particle concentration profile,
\begin{equation}
 \rho(z)=\rho(h_o)e^{-(z-h_o)/z_\rho}
\end{equation}
and a linearly increasing particle velocity profile,
\begin{equation}
 \langle v_x\rangle(z)=\langle v_x\rangle(h_o)+m(z-h_o),
\end{equation}
where $z_\rho\approx10$mm, $\langle v_x\rangle(h_o)\approx1$m/s, and $m\approx70\mathrm{s}^{-1}$ were not varying much with $u_*$. Using these measurements, we obtain,
\begin{widetext}
\begin{equation}
 c_v=\frac{\overline{\langle v_x^2\rangle}}{\overline{\langle v_x\rangle}^2}=\frac{\overline{\langle v_x^2\rangle}}{\overline{\langle v_x\rangle^2}}\times\frac{\overline{\langle v_x\rangle^2}}{\overline{\langle v_x\rangle}^2}=\frac{\overline{\langle v_x^2\rangle}}{\overline{\langle v_x\rangle^2}}\times\frac{\int\limits_{h_o}^\infty\rho\d z\int\limits_{h_o}^\infty\rho\langle v_x\rangle^2\d z}{\left(\int\limits_{h_o}^\infty\rho\langle v_x\rangle\d z\right)^2}=\frac{\overline{\langle v_x^2\rangle}}{\overline{\langle v_x\rangle^2}}\left(1+\left(\frac{mz_\rho}{\langle v_x\rangle(h_o)+mz_\rho}\right)^2\right)\approx1.17\frac{\overline{\langle v_x^2\rangle}}{\overline{\langle v_x\rangle^2}}.
\end{equation}
\end{widetext}
In order to obtain $c_v$, it remains to estimate $\overline{\langle v_x^2\rangle}/\overline{\langle v_x\rangle^2}$. In order to do so, we use measurements of Greeley et al. (Fig. 13 of Ref.~\cite{Greeleyetal96}), who reported a histogram of the horizontal particle velocity of the particles located at a height $z_h=h_o+2cm$, from which we obtain,
\begin{equation}
 \frac{\langle v_x^2\rangle(z_h)}{\langle v_x\rangle^2(z_h)}\approx1.1.
\end{equation}
 Since the shape of the distribution of the horizontal particle velocity does not vary much with the height \cite{RasmussenSorensen08}, we thus estimate $c_v$ as,
\begin{equation}
 c_v\approx1.17\frac{\overline{\langle v_x^2\rangle}}{\overline{\langle v_x\rangle^2}}\approx1.17\frac{\langle v_x^2\rangle(z_h)}{\langle v_x\rangle^2(z_h)}\approx1.3.
\end{equation}

\subsubsection{The parameter $c_U$}
In contrast to $c_v$, the parameter $c_U$ (Eq.~(\ref{cUdef})) significantly influences the functional shape of $L_s$ as a function of $u_{\ast}$. $c_U$ characterizes the significance of the transport-flow feedback for the saturation of the sediment flux. For instance, $c_U=0$ means that the transport-flow feedback does not affect the saturation process since the flow is already saturated ($U=U_s$ from Eq.~(\ref{cUdef})). 

In order to estimate $c_U$, we need to know how the mean fluid speed $U$ behaves as a function of the feedback-reduced bed shear velocity $u_b$ (see Eq.~(\ref{cUdef})). For aeolian sediment transport, the fluid speed is strongly suppressed by the reaction drag forces which the transported grains apply on the wind. The feedback is, in fact, so strong that the mean fluid speed in the transport layer changes only weakly with $u_{\ast}$ \cite{Paehtzetal12a}. In leading-order approximation, the mean fluid speed is thus proportional to $u_b$,
\begin{eqnarray}
 U\approx\frac{U_su_b}{u_{{\mathrm{t}}}}.
\end{eqnarray}
We thus obtain, from Eq.~(\ref{cUdef}), 
\begin{equation}
c_U \approx 1. \label{eq:cu_value}
\end{equation}
We note that a value of $c_U$ close to unity is obtained even if the more complicated dependence of $U$ on $u_b$, obtained from modeling saturated sediment flux \cite{Paehtzetal12a}, is taken into account. Eq.~(\ref{eq:cu_value}) is approximately valid for $u_{\ast} < 4u_{{\mathrm{t}}}$ \cite{Paehtzetal12a}. Beyond this range, turbulence-induced fluctuations of the shear velocity, neglected in the present work, should affect the value of $c_U$.

\subsubsection{The parameter $c_M$}\label{cm_aeolian}
The parameter $c_M$, given by Eq.~(\ref{cMdef})), occurs as a prefactor in Eq.~(\ref{LVfinal2}) and it further affects the functional shape of $L_s(u_{\ast})$ since $c_M$ itself affects the feedback term, $(1-\frac{\d U}{\d V}(V_s))^{-1}$. $c_M$ encodes the relative importance of the respective relaxation processes $M \rightarrow M_s$ and $V \rightarrow V_s$ for the saturation of the sediment flux. There are two extreme cases: $c_M=0$ and $c_M \rightarrow \infty$. The case $c_M = 0$ means that the saturation of $M$ towards $M_s$ is much faster than the saturation of $V$ towards $V_s$, while the opposite situation corresponds to the case $c_M \rightarrow \infty$.  

In order to estimate $c_M$ for the aeolian regime of sediment transport, we first estimate how the function $M(V)$ behaves close to the saturated regime. For this purpose, we make use of the fact that, for aeolian sediment transport, the dominant mechanism which brings grains of the sediment bed into motion is the ejection of bed grains due to impacts of already transported grains, a mechanism known as ``splash'' (see e.g. \cite{KokRenno09,Carneiroetal11,Paehtzetal12a,Carneiroetal13}). It is known that ejection of new grains is mainly due to the impacts of the fastest transported particles, whereas the impacts of slow particles have a negligible effect on the splash process \cite{Andreotti04,KokRenno09,Carneiroetal13}. Indeed, the speed of a fast impacting grain mainly determines the number of ejected grains, but not their ejection velocities, as found in experiments \cite{Beladjineetal07,Ogeretal08}. The ejected particles are typically slow compared to the rebound speed of the impacting particle. In other words, the impact of a fast grain naturally results in two species of particles: a single (fast) rebounding particle and many ejected (slow) particles. 

Using numerical simulations of splash and particle trajectories, Andreotti \cite{Andreotti04} could observe these two distinct species in the characteristics of transported particles. The author noted that the slow species ("reptons") accounts for the majority of transported mass per unit soil area ($M$). Furthermore, the author's analysis suggested that the impact flux of reptons, and thus, in good approximation, the total transported mass $M$, adjusts to changes of the impact flux of the fast species ("saltons") within a distance much shorter than $L_s$. Therefore, it seems reasonable to treat $M$ as locally equilibrated with respect to the impact flux of saltons. The locally equilibrated value $M_{\mathrm{eq}}$ of $M$ is proportional to the number of ejected particles per impact, which in turn is proportional to the impact speed of saltons \cite{KokRenno09} and thus approximately proportional to $V$. A rough estimate of the function $M(V)$ is therefore,
\begin{eqnarray}
 M\approx M_{\mathrm{eq}}\approx\frac{M_sV}{V_s},
\end{eqnarray}
which yields,
\begin{equation}
c_M \approx 1. \label{eq:cM}
\end{equation}

\subsubsection{The Coulomb friction coefficient, $\mu$}
The Coulomb friction coefficient, $\mu$, occurs as a prefactor in Eq.~(\ref{LVfinal2}), and also changes the functional shape of $L_s$ through Eq.~(\ref{vrs2}). $\mu$ can be determined indirectly from measurements of the saturated mass of transported particles $M_s$ as a function of the shear velocity $u_{\ast}$, which fulfills the equation \cite{Paehtzetal12a,Duranetal12},
\begin{eqnarray}
 M_s=\frac{c_a\rho_f}{\mu\tilde g}\cdot \left[u_{\ast}^2-u_{{\mathrm{t}}}^2\right], \label{mudef}
\end{eqnarray}
Note that this equation is Eq.~(\ref{Ms}) with $\tau_{fs}(h_o)=\tau_{\mathrm{t}}$ (Eq.~(\ref{tauftaut})). The value, 
\begin{equation}
{\mu} \approx 1, \label{eq:mu_value}
\end{equation}
was found in a previous work \cite{Paehtzetal12a} through determining $\mu$ indirectly both from experiments as mentioned above and from numerical simulations of aeolian sediment transport in equilibrium.

\subsubsection{The saturated particle velocity $V_s$}
The saturated particle velocity $V_s$ is dominantly controlling the dependence of $L_s$ on $u_{\ast}$ in Eq.~(\ref{LVfinal2}). Since the dependence of $V_s$ on $u_{\ast}$ is rather weak for aeolian sediment transport \cite{Paehtzetal12a}, the saturation length $L_s$ will not change much with $u_{\ast}$. Here we use an expression for $V_s$ which has been obtained in a recent work \cite{Paehtzetal12a}, since the values of saturated sediment flux obtained using this equation produced excellent quantitative agreement with measurements \cite{Paehtzetal12a}. The expression for $V_s$ reads \cite{Paehtzetal12a},
\begin{equation}
 V_s=V_{\mathrm{t}}+\frac{3u_{{\mathrm{t}}}}{2\kappa}\cdot\ln\left(\frac{V_s}{V_{\mathrm{t}}}\right)+\frac{u_{\ast}}{\kappa}\cdot F_\gamma\!\left(\frac{u_{{\mathrm{t}}}}{u_{\ast}}\right), \label{Vsaeolian}
\end{equation}
where $V_{\mathrm{t}}$ and $F_\gamma(x)$ are given by the equations,
\begin{eqnarray}
& & V_{\mathrm{t}}=\frac{V_o+\eta V_{rs}}{1-\eta}\quad\mathrm{with}\quad V_o=16.2\sqrt{\tilde gd+\frac{6\zeta}{\pi\rho_pd}}, \label{eq:V_t} \\
& & F_\gamma(x)=(1-x)\cdot\ln(1.78\gamma)+0.5\cdot(1-x^2)\cdot\mathrm{E_1}\!(\gamma)  \nonumber \\
& & \ \ \ \ \ \ \ \ +1.154\cdot(1+x\ln x)\cdot{(1-x)^{2.56}}. \label{F} 
\end{eqnarray}
In these equations, $\mathrm{E_1}(x)$ is the exponential integral function, $\kappa=0.4$ is the von K\'arm\'an constant, and $\zeta=5\times10^{-4}N/m$ is a dimensional parameter encoding the influence of cohesion, while $\eta=0.1$ and $\gamma=0.17$ are empirically determined parameters \cite{Paehtzetal12a}. The saturated particle velocity $V_s$ for transport in the aeolian regime can be obtained by iteratively solving Eq.~(\ref{Vsaeolian}) for $V_s$ and using the expressions for $V_{\mathrm{t}}$ and $F_{\gamma}(x)$ given by Eqs.~(\ref{eq:V_t}) and (\ref{F}), respectively.

\subsubsection{The threshold shear velocity $u_{{\mathrm{t}}}$}
We calculate the threshold shear velocity $u_{{\mathrm{t}}}$ by using the following equation, which has been obtained from an analytical model for aeolian sediment transport in equilibrium \cite{Paehtzetal12a},
\begin{equation}
 u_{{\mathrm{t}}}=\frac{\kappa\cdot(V_{rs}+V_o)}{(1-\eta)\cdot\ln{\left({z_{mt}}/{z_o}\right)}}, \label{ut} \\
\end{equation}
where $z_{mt}$ is given by the following equation \cite{Paehtzetal12a},
\begin{eqnarray}
 z_{mt}&=&\frac{\beta\gamma V_{rs}^\frac{1}{2}V_{\mathrm{t}}^\frac{3}{2}}{\mu\tilde g}. \label{zm2}
\end{eqnarray}
In the equation above, $\beta=0.095$ is an empirically determined parameter \cite{Paehtzetal12a}, while $z_o$, which is the surface roughness of the quiescent sediment bed, is given by the equation \cite{ChengChiew98,Paehtzetal12a},
\begin{eqnarray}
& & z_o=d\exp(-\kappa B),\quad\mathrm{with,} \label{zo} \\
& & B=8.5+(2.5\ln R_p-3)\exp\left[-0.11(\ln R_p)^{2.5}\right], \nonumber
\end{eqnarray}
where $R_p=\frac{u_{{\mathrm{t}}}d}{\nu}$. We note that the ratio between $u_{{\mathrm{t}}}$ (which is the threshold for sustained transport) and the fluid threshold $u_{\mathrm{ft}}$ required to initiate transport in the aeolian regime depends strongly on the environmental conditions. Eq.~(\ref{ut}) yields $u_{{\mathrm{t}}} \approx 0.8u_{\mathrm{ft}}$ for Earth conditions, which is in agreement with measurements \cite{Bagnold41}. However, the ratio $u_{{\mathrm{t}}}/u_{\mathrm{ft}}$ under Martian conditions can be as small as $10\%$, as also found from numerical simulations \cite{Kok10a,Koketal12}. Indeed, Eq.~(\ref{ut}), which was obtained from the same theoretical work leading to Eq.~(\ref{Vsaeolian}), has been validated by comparing its prediction with outcomes of numerical simulations \cite{KokRenno09} under a wide range of fluid-to-sediment density ratio and particle diameter, thereby leading to excellent quantitative agreement (see Fig. 13b of Ref.~\cite{Paehtzetal12a}).

\subsection{The saturation length of subaqueous sediment transport}\label{Ls_subaqueous}
In this section, we provide expressions for the parameters $c_v$, $c_U$ and $c_M$, as well as for the Coulomb friction coefficient, $\mu$, the saturated particle velocity, $V_s$, and the threshold shear velocity, $u_{{\mathrm{t}}}$, for transport in the subaqueous regime. We remark that we estimate these quantities only in a rough manner, consistent with the large scatter (factor $2-4$) of the experimental data.

\subsubsection{The parameter $c_v$}\label{cv_subaqueous}
In this section, we reiterate some of the results we obtained in Section~A2 of the supplementary material of Ref.~\cite{Paehtzetal13}. We can estimate $c_v$ for transport in the subaqueous regime from measurements of the distribution $P_v(v_x)$ of horizontal velocities $v_x$ in subaqueous sediment transport in equilibrium. Such measurements were undertaken by Lajeunesse et al. in experiments of sediment transport under water using particles of average diameter $d=2.24\,$mm and relative shear velocity $u_{\ast}/u_{{\mathrm{t}}}=2.1$ \cite{Lajeunesseetal10}. In these experiments, particles were considered as being transported if they had a velocity larger than a certain cut-off value, $v_c$ \cite{Lajeunesseetal10}. The distribution of horizontal velocities for these transported particles was fitted using an exponential distribution,
\begin{equation}
 P_v(v_x)=\frac{1}{V_f}{\mbox{exp}}{\left[{-\frac{v_x-v_c}{V_f}}\right]},
\end{equation}
where $V_f\approx 110\,$mm$/$s. By using this distribution, we can compute $c_v$ as,
\begin{equation}
 c_v=\frac{\int\limits_{v_c}^\infty v_x^2P_v(v_x)\d v_x}{\left(\int\limits_{v_c}^\infty v_xP_v(v_x)\d v_x\right)^2}=\frac{1+\left(1+\frac{v_c}{V_f}\right)^2}{\left(1+\frac{v_c}{V_f}\right)^2}. \label{estcv}
\end{equation}
Lajeunesse et al. did not report specific values of $v_c$ corresponding to specific measurements \cite{Lajeunesseetal10}. Instead they mentioned that $v_c$ lies within the range between $10\,$mm$/$s and $30\,$mm$/$s, depending on the water flow rate. Since $d=2.24\,$mm and $u_{\ast}/u_{{\mathrm{t}}}=2.1$ (which are the values reported for the measurement of $P_v(v_x)$) correspond to intermediate values for $d$ and $u_{\ast}/u_{{\mathrm{t}}}$ investigated in the experiments \cite{Lajeunesseetal10}, we use the intermediate value $v_c=20\,$mm$/$s as an approximate estimate for the average cut-off velocity. Using this estimate for $v_c$, Eq.(\ref{estcv}) yields,
\begin{equation}
c_v\approx1.7, \label{eq:cv_value_subaqueous}
\end{equation}
for transport in the subaqueous regime.

\subsubsection{The parameter $c_U$}\label{cu_subaqueous}
In contrast to the aeolian regime, the suppression of the fluid flow due to the sediment transport in the subaqueous sediment transport is weak \cite{Duranetal12}. The mean fluid speed $U$ is thus mainly a function of the shear velocity $u_{\ast}$ and the dependence of $U$ on $u_b$ and thus on $V$ is negligible. By neglecting this dependence, we obtain,
\begin{equation}
c_U \approx 0, \label{eq:cu_value_subaqueous}
\end{equation}
which is consequence of Eq.~(\ref{cUdef}) with $\d U / \d u_b \approx 0$.

\subsubsection{The parameter $c_M$}

In order to estimate $c_M$ for subaqueous sediment transport, we use evidence provided by the recent numerical study of Dur\'an et al. \cite{Duranetal12}. As mentioned before, these authors simulated the dynamics of both the transported particles and the sediment bed at the single particle scale. Dur\'an et al. \cite{Duranetal12} found that, during flux saturation in subaqueous sediment transport, $M$ changes within a time scale which is more than one order of magnitude larger than the time scale in which $Q$ changes. This observation can be mathematically expressed as,
\begin{eqnarray}
 \left|V\frac{\d M}{\d t}\right|\ll\left|\frac{\d Q}{\d t}\right|,
\end{eqnarray}
and thus,
\begin{eqnarray}
 \left|V\frac{\d M}{\d t}\right|\ll\left|M\frac{\d V}{\d t}\right|. \label{dMvsdV}
\end{eqnarray}
Eq.~(\ref{dMvsdV}) further implies that,
\begin{eqnarray}
 \left|\frac{V}{M}\frac{\d M}{\d V}\right|\ll1,
\end{eqnarray}
and thus
\begin{eqnarray}
 |c_M|\ll1,
\end{eqnarray}
where we used the definition of $c_M$, which is given by Eq.~(\ref{cMdef}). Hence, we estimate $c_M$ as,
\begin{equation} 
c_M \approx 0. \label{eq:cm_value_subaqueous}
\end{equation}
However, we note that our model predictions are consistent with experiments even if we assume a coupling of $M$ to $V$ which is as strong as in the aeolian regime --- that is, even by assuming $c_M=1$ and thus increasing $L_{\mathrm{s}}$ by a factor of $1.5$ as compared to the value obtained with $c_M=0$ \cite{Paehtzetal13}. This means that the saturation length in the subaqueous regime is not very sensitive to the value of $c_M$ within the range between $0$ and $1$ (whereas the latter value corresponds to sediment transport in the aeolian regime).

\subsubsection{The parameter $\mu$}
In this section, we reiterate some of the results we obtained in Section~B of the supplementary material of Ref.~\cite{Paehtzetal13}. As obtained in experiments on subaqueous sediment transport in equilibrium, the average mass flux $M_s$ approximately follows the expression \cite{Lajeunesseetal10},
\begin{eqnarray}
 M_s=\frac{c_a}{0.415\tilde g}\cdot\left[\tau-\tau_{\mathrm{t}}\right].
\end{eqnarray}
By comparing this equation with Eq.~(\ref{Ms}) with $\tau_{fo}=\tau_{\mathrm{t}}$ and $c_a=1.19$ (see section \ref{equations_for_Ls}), we obtain,
\begin{equation}
\mu \approx 0.493. \label{eq:mu_subaqueous_value}
\end{equation}
for sediment transport in the subaqueous regime.

\subsubsection{The saturated particle velocity $V_s$}
\begin{figure}
 \begin{center}
  \includegraphics[scale=0.28]{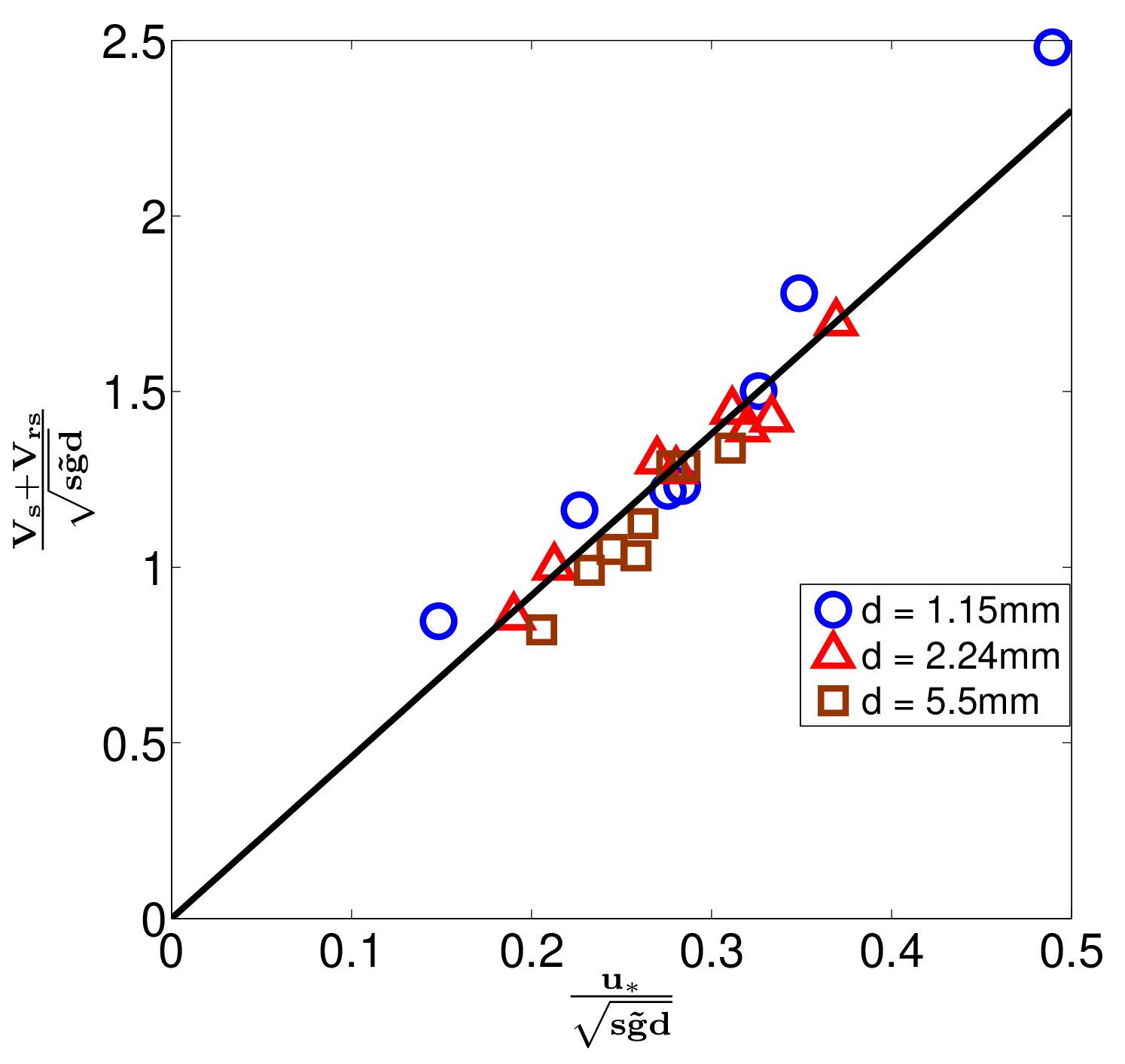}
 \end{center}
 \caption{Average value of the dimensionless fluid speed $\frac{U_s}{s\tilde gd}=\frac{V_s+V_{rs}}{s\tilde gd}$, as a function of the dimensionless shear velocity $\frac{u_{\ast}}{s\tilde gd}$. For the symbols, the average dimensionless particle speed $\frac{V_s}{s\tilde gd}$ was obtained from measurements \cite{Lajeunesseetal10}, while we computed $V_{rs}$ using Eq.~(\ref{vrs2}) with $\mu=0.493$. The black solid line corresponds to the best fit to the experimental data using Eq.~(\ref{VsVrs}), which yields $a\approx4.6$.}
 \label{aplot}
\end{figure}

It has been verified in a large number of experimental studies \cite{Bagnold56,Bagnold73,FernandezLuqueBeek76,VanRijn84,HuHui96,Seminaraetal02,Lajeunesseetal10}, that the equilibrium particle velocity in the subaqueous regime of transport approximately follows the expression,
\begin{eqnarray}
 V_s=au_{\ast}-V_{rs}, \label{VsVrs}
\end{eqnarray}
where $a$ is a dimensionless number. We note that the above expression is consequence of the equation, $V_s = U_s - V_{rs}$, where $U_s$ is taken proportional to $u_{\ast}$. In order to obtain $V_s$ for sediment transport in the subaqueous regime using Eq.~(\ref{VsVrs}), we calculate $V_{rs}$ using Eq.~(\ref{vrs2}) and use the value $a\approx4.6$, which we have obtained by comparing the prediction of Eq.~(\ref{VsVrs}) with measurements of $V_s$ as a function of $u_{\ast}$ from experiments on subaqueous sediment transport in equilibrium \cite{Lajeunesseetal10} (see \figurename~\ref{aplot}).

\subsubsection{The threshold shear velocity $u_{{\mathrm{t}}}$}
The threshold velocity for sustained sediment transport, $u_{{\mathrm{t}}}$, in the subaqueous regime is computed by using the equation,
\begin{equation}
u_{{\mathrm{t}}} = \sqrt{\Theta_{\mathrm{t}}s\tilde gd}, \label{Shields}
\end{equation}
where the threshold Shields parameter $\Theta_{\mathrm{t}}$ is obtained through an empirical fit to the Shields diagram \cite{Paphitis01}. The resulting expression for $\Theta_{\mathrm{t}}$ reads \cite{Paphitis01},
\begin{equation}
 \Theta_{\mathrm{t}}=\frac{0.273}{1+1.2D_*}+0.046\cdot\left(1-0.576e^{-0.02D_*}\right),
\end{equation}
where, $D_*=d\sqrt[3]{s\tilde g/\nu^2}$.

\section{Dependence of the saturation length on particle size and fluid shear velocity}\label{Discussion}

In order to understand the morphodynamics of sediment landscapes under water and on planetary surfaces, it is important to understand the behavior of the flux saturation length as a function of the relevant attributes of sediment and fluid. In particular, the size of planetary dunes can serve as a proxy for the saturation length of extraterrestrial dune fields, which can be used to infer the local fluid shear velocity ($u_{\ast}$) and average size ($d$) of the constituent sediment \cite{Kroyetal02a,Partelietal07,ParteliHerrmann07b,Koketal12}. In fact, the dependence of $L_s$ on $u_{\ast}$ has been subject of intense debate in previous theoretical works \cite{Sauermannetal01,Hersenetal02,Partelietal07,Andreottietal10,Fourriereetal10}. It is therefore useful to perform in this Section a systematic study of the saturation length as a function of these two relevant parameters under different environmental conditions.

\subsection{The saturation length as a function of the fluid shear velocity, $u_{\ast}$}

\begin{figure}
 \begin{center}
  \includegraphics[scale=0.28]{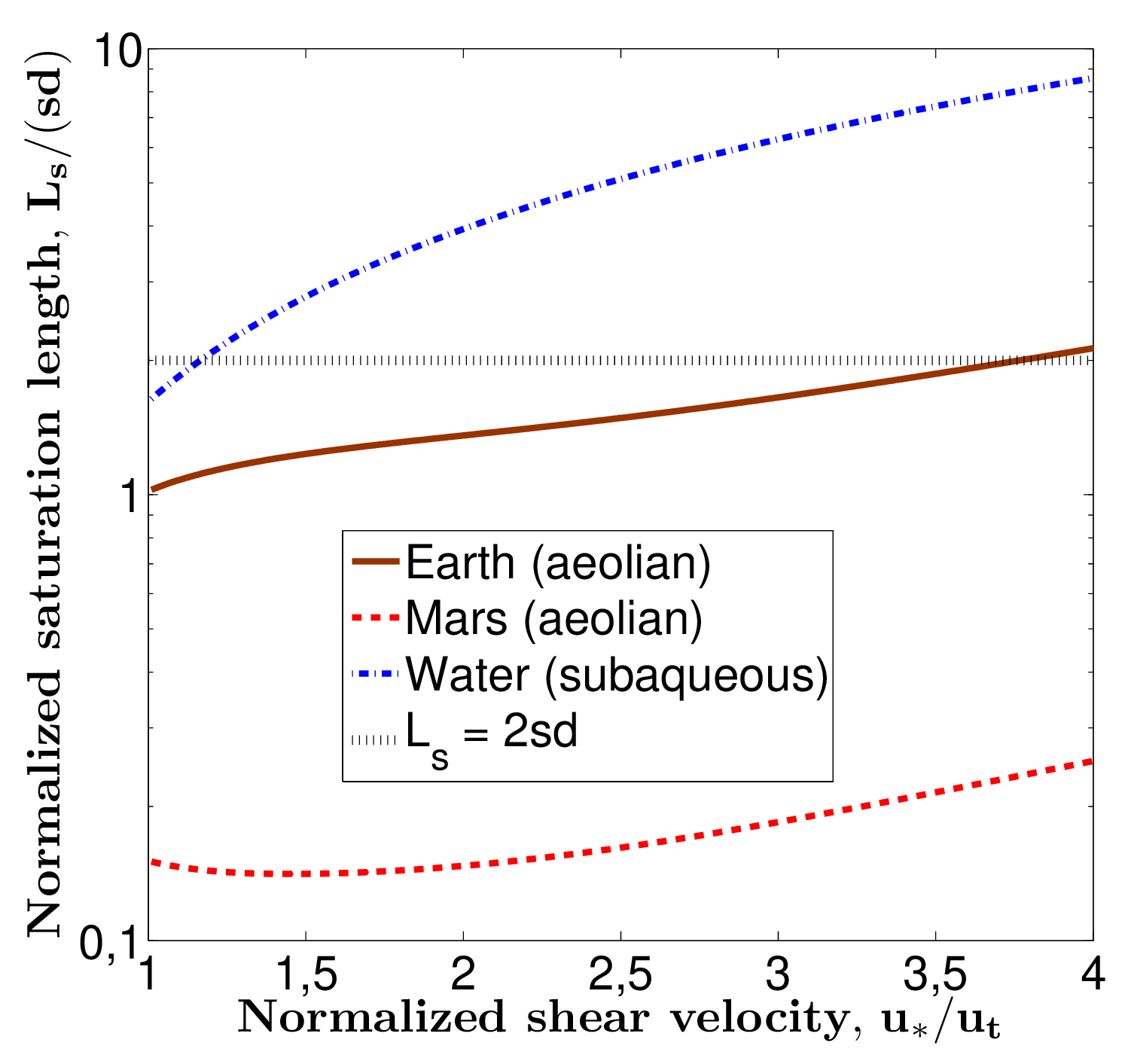}
 \end{center}
 \caption{Dimensionless saturation length $L_s/(sd)$ versus dimensionless shear velocity $u_*/u_{\mathrm{t}}$ for particles with mean diameter $d=250\mathrm{\mu m}$, computed using Eq.~(\ref{LVfinal2}). The brown, solid line corresponds to aeolian sediment transport on Earth ($\rho_p=2650\mathrm{kg/m^3}$, $\rho_w=1.2\mathrm{kg/m^3}$, $g=9.81\mathrm{m/s^2}$, $\nu=1.5\times10^{-5}\mathrm{m^2/s}$), the red, dashed line corresponds to aeolian sediment transport on Mars ($\rho_p=3000\mathrm{kg/m^3}$, $\rho_w=0.0185\mathrm{kg/m^3}$, $g=3.71\mathrm{m/s^2}$, $\nu=6.4\times10^{-4}\mathrm{m^2/s}$), and the blue, dash-dotted line corresponds to subaqueous sediment transport under water ($\rho_p=2650\mathrm{kg/m^3}$, $\rho_w=1000\mathrm{kg/m^3}$, $g=9.81\mathrm{m/s^2}$, $\nu=10^{-6}\mathrm{m^2/s}$). Moreover, the black, dotted line shows $L_s=2sd$ as proposed by Refs. \cite{Hersenetal02,ClaudinAndreotti06,Andreottietal10,Fourriereetal10}.}
 \label{Ls_u}
\end{figure}

\figurename~\ref{Ls_u} shows the dependence of $L_s/(sd)$ on $u_*/u_{\mathrm{t}}$ for aeolian sediment transport on Earth (brown solid line) and Mars (red dashed line) and for sediment transport under water (blue dash-dotted line) computed using Eq.~(\ref{LVfinal2}) for particles with mean diameter $d=250\mathrm{\mu m}$.
The behavior of $L_s$ with $u_{\ast}$ as predicted from Eq.~(\ref{LVfinal2}) is in clear contrast to the scaling relation $L_s\approx2sd$ proposed in previous works \cite{Hersenetal02,ClaudinAndreotti06,Andreottietal10,Fourriereetal10}. This approximate scaling, which includes no dependence of $L_s$ on $u_{\ast}$, was obtained by assuming that the acceleration of transported particles due to fluid drag is the dominant relaxation mechanism, and by neglecting the entrainment of sediment bed particles due to fluid lift as well as the entrainment of sediment bed particles and the deceleration of transported particles resulting from grain-bed collisions. In our more comprehensive model for saturation of sediment flux, however, all these aforementioned relaxation processes are taken into account. There are two main reasons for the disparity in the behavior of $L_{s}$ with $u_{\ast}$ as observed in our model and in the model of Refs.~\cite{Hersenetal02,ClaudinAndreotti06,Andreottietal10,Fourriereetal10}.

First, our expression involves a significant dependence of $L_s$ on $u_*/u_{\mathrm{t}}$ due to the dependence of $L_s$ on the average particle velocity $V_s$ and the feedback term $K$ (see Eq.~(\ref{Lsfluid})), both of which are functions of $u_*/u_{\mathrm{t}}$. In particular, for the subaqueous regime, $V_s$ is a strongly increasing function of $u_*/u_{\mathrm{t}}$ thus explaining the strong increase of $L_s$ with $u_*/u_{\mathrm{t}}$ in this regime. Furthermore, we see in Fig.~\ref{Ls_u} that in our model the dependence of $L_s$ on $u_{\ast}$ in the aeolian regime is small, but not negligible as suggested in the model of Refs.~\cite{Hersenetal02,ClaudinAndreotti06,Andreottietal10,Fourriereetal10}. Indeed, the dependence of $L_s$ on $V_s$ and $K$ is a consequence of considering grain-bed collisions and the transport-flow feedback, respectively, for the saturation of the sediment mass flux $Q$ --- both neglected in the models of Refs.~\cite{Hersenetal02,ClaudinAndreotti06,Andreottietal10,Fourriereetal10}. Second, in contrast to the models proposed in these works, our model considers the dependence of the drag coefficient $C_d$ on the particle Reynolds number, $R_{ep}=V_{rs}d/\nu$ (see Eq.~\ref{drag}). As can be seen in \figurename~\ref{Ls_u}, the difference between the particle Reynolds numbers on Earth ($R_{ep}\approx30$) and Mars ($R_{ep}\approx1$) results in an order of magnitude difference between $L_s/(sd)$ on these two planetary bodies, which occurs because the normalized particle velocity $V_s/\sqrt{s\tilde gd}$ increases strongly with $R_{ep}$ when $R_{ep}$ is of order unity \cite{Paehtzetal12a}.

It can also be seen in \figurename~\ref{Ls_u} that the saturation length $L_s$ on Mars decreases with $u_*$ when $u_*$ is sufficiently close to $u_{\mathrm{t}}$, even though $V_s$ increases with $u_{\ast}$ in this regime. This surprising behavior is a consequence of the feedback term $K$, which, for Mars conditions, decreases with $u_*$ sufficiently close to $u_{\mathrm{t}}$ and thus overcompensates the tendency of $L_s$ to increase with $V_s$. In contrast, for Earth conditions, the feedback term $K$ increases with $u_*$ close to $u_{\mathrm{t}}$. This qualitative difference in the change of $K$ with $u_*$ between Earth and Mars conditions can be understood by noting that,
\begin{equation}
 \frac{V_s}{c_MFV_{rs}}<1 \label{diffMarsEarth1}
\end{equation}
for Mars conditions with sufficiently small $u_*/u_{\mathrm{t}}$, while for Earth conditions (and for Mars conditions with sufficiently large $u_*/u_{\mathrm{t}}$) the following relation holds,
\begin{equation}
 \frac{V_s}{c_MFV_{rs}}>1. \label{diffMarsEarth2}
\end{equation}

The physical origin for the difference in the behavior of $K$ with $u_{\ast}$ mentioned above lies in the mechanics of the reduction of the fluid speed due to sediment transport (see Section~\ref{derivation_of_Ls}). The fluid velocity in the transport layer ($U$) decreases with the average drag force applied by the fluid onto the transport layer ($\frac{3M}{4sdc_a}C_d(V_r)V_r^2$). This average drag force in turn is proportional to both the mass density $M$ of transported particles and to the acceleration term $C_d(V_r)V_r^2$. Recalling that $c_M\approx1$ in the aeolian regime, we have that, close to saturation, both $M$ and $V$ are smaller (if $Q<Q_s$) or larger (if $Q>Q_s$) than their respective saturated values, $M_s$ and $V_s$. Hence, if $M<M_s$ ($M>M_s$), it follows that $C_d(V_r)V_r^2>C_d(V_{rs})V_{rs}^2$ ($C_d(V_r)V_r^2<C_d(V_{rs})V_{rs}^2$), which means that the mass density and the acceleration term deviate from their saturated values in opposite directions. The average drag force applied by the fluid onto the transport layer ($\frac{3M}{4sdc_a}C_d(V_r)V_r^2$) thus can be both larger (under Mars conditions for sufficiently small $u_*/u_{\mathrm{t}}$) or smaller (under both terrestrial and Martian conditions for sufficiently large $u_*/u_{\mathrm{t}}$) than its saturated value. Consequently, the fluid velocity in the transport layer ($U$) can be smaller or larger than its saturated value $U_s$, depending on whether Eq.~(\ref{diffMarsEarth2}) or Eq.~(\ref{diffMarsEarth1}), respectively, is fulfilled. 
Further, if $U>U_s$ (Eq.~(\ref{diffMarsEarth1})), then $K<1$ and thus the saturation length $L_s$ decreases in comparison to the situation in which the saturation of feedback is neglected ($K=1$). In contrast, if $U<U_s$ (Eq.~(\ref{diffMarsEarth2})), $K>1$ and thus the saturation length $L_s$ increases in comparison to $K=1$. The deviation of $K$ from $K=1$ becomes stronger with increasing $u_*/u_{\mathrm{t}}$ because the effect of the sediment transport on the fluid velocity increases with $u_*/u_{\mathrm{t}}$. This explains why the feedback term $K$ can both increase or decrease with $u_*/u_{\mathrm{t}}$.

\subsection{Dependence of the saturation length on the average particle size ($d$) }

\begin{figure*}
 \begin{center}
  \includegraphics[scale=0.35]{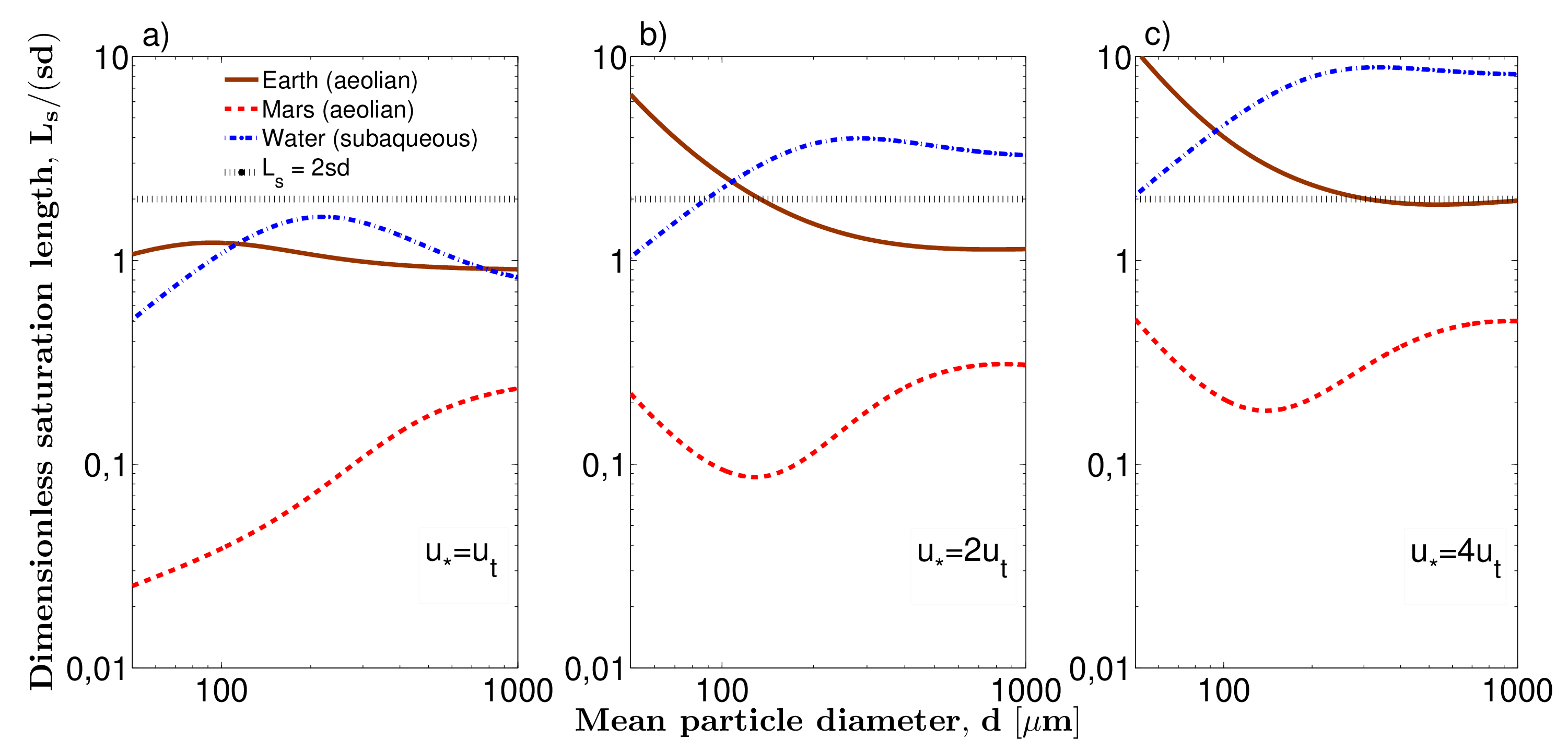}
 \end{center}
 \caption{Normalized saturation length $L_s/(sd)$ versus mean particle diameter $d$ for three shear velocities, namely, $u_*=u_{\mathrm{t}}$, $u_*=2u_{\mathrm{t}}$, and $u_*=4u_{\mathrm{t}}$, computed using Eq.~(\ref{LVfinal2}). The brown, solid line corresponds to aeolian sediment transport on Earth ($\rho_p=2650\mathrm{kg/m^3}$, $\rho_w=1.2\mathrm{kg/m^3}$, $g=9.81\mathrm{m/s^2}$, $\nu=1.5\times10^{-5}\mathrm{m^2/s}$), the red, dashed line corresponds to aeolian sediment transport on Mars ($\rho_p=3000\mathrm{kg/m^3}$, $\rho_w=0.0185\mathrm{kg/m^3}$, $g=3.71\mathrm{m/s^2}$, $\nu=6.4\times10^{-4}\mathrm{m^2/s}$), and the blue, dash-dotted line corresponds to subaqueous sediment transport under water ($\rho_p=2650\mathrm{kg/m^3}$, $\rho_w=1000\mathrm{kg/m^3}$, $g=9.81\mathrm{m/s^2}$, $\nu=10^{-6}\mathrm{m^2/s}$). Moreover, the black, dotted line shows $L_s=2sd$ as proposed by Refs. \cite{Hersenetal02,ClaudinAndreotti06,Andreottietal10,Fourriereetal10}.}
 \label{Ls_d}
\end{figure*}

Fig.~\ref{Ls_d} shows the dependence of $L_s/(sd)$ on $d$ for aeolian sediment transport on Earth (brown solid line) and Mars (red dashed line), and for subaqueous sediment transport (blue dash-dotted line), computed using Eq.~(\ref{LVfinal2}) for different values of the fluid shear velocity, namely, $u_*=u_{\mathrm{t}}$, $u_*=2u_{\mathrm{t}}$, and $u_*=4u_{\mathrm{t}}$.
It can be seen that the rescaled saturation length $L_s/(sd)$ displays a complex behavior with $d$. The dependence of $L_s/(sd)$ on $d$ is controlled by two main factors: the dependence of the drag coefficient and of the Shields parameter on the particle Reynolds number ($R_{ep}$), and the dependence of the saturation of the sediment flux on the transport-flow feedback. The significance of each one of these factors for the dependence of the saturation length on the average grain diameter depends on the transport regime, as we will discuss in the next paragraphs.

{\em{Subaqueous regime of transport}} --- The change in the transport-flow feedback due to the saturation of the flux is negligible in the subaqueous regime, since in this regime $c_U\approx0$ and thus $K\approx1$. This behavior explains why all curves $L_s/(sd)$ versus $d$ in Figs.~\ref{Ls_d}a$-$c corresponding to transport under water display the same qualitative behavior independent of $u_{\ast}$. In each of these curves, $L_s/(sd)$ first increases with $d$, then reaches a maximum, and finally approaches a constant value. The origin of this behavior is that $L_s$ is proportional to $V_{rs}V_sF$ (cf. Eq.~(\ref{LVfinal2})), and thus its dependence on the grain size is determined by the behavior of $V_{rs}V_sF$ with $d$. For sufficiently large particle diameters $d$ (large $R_{ep}$), both the drag coefficient $C_d$ and the Shields parameter $\Theta_{\mathrm{t}}$ are approximately independent of $d$, and thus both $V_{rs}/\sqrt{s\tilde gd}$ and $V_s/\sqrt{s\tilde gd}$ are roughly independent of $d$ (see Eq.~(\ref{vrs2}), as well as Eqs.~(\ref{VsVrs}) and (\ref{Shields}) for constant $u_*/u_{\mathrm{t}}$), while $F\approx0.5$ (see Eq.~(\ref{Cd_Fx})). Hence, $L_s/(sd)$ is nearly independent of $d$ for sufficiently large particle diameters in the subaqueous regime. However, for smaller particle diameters $d$ (smaller $R_{ep}$), both $C_d$ and $\Theta_{\mathrm{t}}$, and thus $V_{rs}/\sqrt{s\tilde gd}$, $V_s/\sqrt{s\tilde gd}$ and $F$, incorporate a dependence on $d$. In this regime, $V_{rs}/\sqrt{s\tilde gd}$ increases with $d$, whereas both $V_s/\sqrt{s\tilde gd}$ and $F$ decrease with $d$. The increase of $V_{rs}/\sqrt{s\tilde gd}$ with $d$ thereby dominates the behavior of $L_s/(sd)$ for small particle diameters, while the decrease of $V_s/\sqrt{s\tilde gd}$ and $F$ with $d$ dominates the behavior of the saturation length with $d$ for large particle diameters. 

{\em{Aeolian regime under terrestrial conditions}} --- The same qualitative behavior of $L_s/(sd)$ with $d$ observed in the subaqueous regime occurs for aeolian sediment transport on Earth at shear velocities close to the threshold, as can be seen in Fig.~\ref{Ls_d}a. Two main factors dictate the observed dependence of $L_s/(sd)$ on $d$ for aeolian transport under terrestrial conditions. First, close to the threshold, the fluid velocity is almost undisturbed by the particle transport, since only a few particles are transported. Hence, the feedback term $K$ associated with the aeolian regime of transport under terrestrial conditions when $u_{\ast}$ is close to threshold for sustained transport, $u_{\mathrm{t}}$, is close to unity, as it is in the subaqueous regime. Second, $V_{rs}/\sqrt{s\tilde gd}$, $V_s/\sqrt{s\tilde gd}$ and $F$ behave qualitatively in the same manner with $d$ as they do in the subaqueous regime. Therefore, also $L_s/(sd)$ for aeolian transport under terrestrial conditions with $u_{\ast}$ close to $u_{\mathrm{t}}$ depends on $d$ in the same manner as it does in the subaqueous regime. In contrast, for large shear velocities, the qualitative behavior of $L_s/(sd)$ with $d$ observed for aeolian sediment transport on Earth is qualitatively different from the one observed in the subaqueous regime (see Figs.~\ref{Ls_d}b$-$c). This is because the saturation of the transport-flow feedback for aeolian sediment transport plays a relevant role for large $u_{\ast}$. In this regime, the approximation $K\approx1$ is not valid anymore, instead the feedback term $K$ follows Eq.~(\ref{Kasym}). Consequently, $L_s$ is proportional to $V_s^2$ and not to $V_{rs}V_sF$. Since $V_s/\sqrt{s\tilde gd}$ decreases with $d$ before it reaches an approximately constant value, so does $L_s/(sd)$ as can be seen in Figs.~\ref{Ls_d}b$-$c.

{\em{Aeolian regime under Martian conditions}} --- For aeolian transport under Martian conditions, $L_s/(sd)$ shows a qualitative behavior with $d$ that is different from the one of aeolian transport under terrestrial conditions. The origin of this discrepancy is a different qualitative behavior of the normalized average particle velocity $V_s/\sqrt{s\tilde gd}$, as we explain in the following. In aeolian sediment transport, the average velocity of particles impacting and leaving the sediment bed ($V_o$) is nearly independent of properties of the fluid \cite{Koketal12,Bourkeetal10,Paehtzetal12a,Creysselsetal09,Duranetal11,RasmussenSorensen08,KokRenno09,Andreotti04,Kok10a}. Rather, $V_o$ is largely controlled by the characteristics of the sediment bed, for instance by cohesive interparticle forces. These forces increase in importance with decreasing particle size \cite{Koketal12}. Due to these forces, $V_o/\sqrt{\tilde gd}$ increases with decreasing $d$ (see Eq.~(\ref{eq:V_t})). At the same time, as already mentioned, $V_{rs}/\sqrt{s\tilde gd}$ increases with $d$. Both $V_o/\sqrt{s\tilde gd}$ and $V_{rs}/\sqrt{s\tilde gd}$ control the normalized average particle velocity $V_s/\sqrt{s\tilde gd}$ (see Eqs.~(\ref{Vsaeolian}) and (\ref{eq:V_t})).

The qualitative difference between aeolian sediment transport on Earth and Mars is due to $V_o/V_{rs}$ being approximately $2.5$ times larger on Earth than on Mars. This difference in the scaling of $V_o/V_{rs}$ implies that the decreasing trend of $V_o/\sqrt{s\tilde gd}$ with $d$ has a smaller effect on the saturation length for Martian conditions than it does for $L_s$ under terrestrial conditions. Indeed, while the value of $V_s/\sqrt{s\tilde gd}$ of aeolian transport under terrestrial conditions decreases with $d$ (due to the decrease of $V_o/\sqrt{s\tilde gd}$ with $d$), the Martian value of $V_s/\sqrt{s\tilde gd}$ {\em{increases}} with $d$. This behavior, which is opposite to the one of terrestrial transport, is due to the increase of $V_{rs}/\sqrt{s\tilde gd}$ with $d$ --- except for small particle diameters for which $V_{rs}/\sqrt{s\tilde gd}$ becomes much smaller than $V_o/\sqrt{s\tilde gd}$. 

The complex behavior of the saturation length $L_s$ plotted in Figs.~\ref{Ls_u} and \ref{Ls_d} suggests that both the entrainment of sediment bed particles by fluid lift and grain-bed collisions, and the momentum change of saltating particles due to drag and grain-bed collisions have a considerable influence on the saturation of the sediment flux. Thus, these relaxation mechanisms of the sediment flux probably cannot be neglected as done in previous studies \cite{Sauermannetal01,Hersenetal02,Charru06,ClaudinAndreotti06,Andreottietal10,Fourriereetal10}. Moreover, the relaxation of $U$, which has also been neglected in previous studies, seems to play a significant role in the saturation length of aeolian sediment transport, for which the feedback term $K$ (Eq.~\ref{Lsfluid}) does not vanish since $c_U\approx c_M\approx1$. In fact, for sufficiently large $u_*$, the feedback term $K$ follows Eq.~(\ref{Kasym}), which means that $L_s$ scales with $V_s^2$ instead of scaling with $V_{rs}V_sF$ for small $u_*$. Hence, our study suggests that the saturation of $U$ cannot be neglected for aeolian sediment transport.


\section{Conclusions}\label{Conclusions}

In conclusion, we have presented a model for flux saturation in sediment transport which, for the first time, accounts for both relevant relaxation processes of sediment flux identified in previous works --- namely, the saturation of the mass density of transported particles and the relaxation of particle velocities --- as well as for the different types of sediment entrainment prevailing under different environmental conditions. Furthermore, our model accounts for the saturation transient of the fluid velocity within the transport layer, which is associated with the saturation of the transport-flow feedback inherent to the interaction between the fluid and the particles in transport. The main outcome of our analytical treatment is a closed-expression for the saturation length of sediment transport, $L_s$ (Eq.~(\ref{LVfinal2})), which can be used to calculate $L_s$ under different environmental conditions corresponding to both subaqueous and aeolian regimes of sediment transport. In particular, $L_s$ predicted from our equation is a complex function of the grain diameter, $d$, and of the fluid shear velocity, $u_{\ast}$. This behavior is in contrast with the scaling of $L_s$ with $sd$ \cite{Hersenetal02,ClaudinAndreotti06,Andreottietal10,Fourriereetal10}, which was obtained from a simplified model that considers only the relaxation of particle velocity and thus neglects the dependence of $L_s$ on $u_{\ast}$ observed in experiments \cite{FranklinCharru11}. 

While the purpose of the present work was to introduce our theoretical model for flux saturation and to present the analytical derivation of our universal equation for the saturation length, in a separate work \cite{Paehtzetal13} we show that this equation consistently predicts the saturation length in different physical environments. Indeed, our saturation length equation is in good quantitative agreement with direct measurements of $L_s$ in a wind tunnel, as well as with indirect estimates of $L_s$ from the size of subaqueous ripples and dunes on Earth, Mars and Venus.

In future studies, our equation can be used to predict the scale of dunes under different extraterrestrial environments \cite{Paehtzetal13} or to infer attributes of sediment and flow in planetary dune fields from the minimal size of barchan dunes or from the wavelength of ``elementary'' dunes emerging on dense sand beds. Moreover, our equation can be used to calculate the saturation length in the morphodynamic dune model of Ref.~\cite{Kroyetal02a}, which couples a continuum model for sediment transport with an analytical model for the average turbulent fluid shear stress over mildly-sloped topographies \cite{JacksonHunt75,Huntetal88}. Our equation can further improve morphodynamic models in hydraulic engineering applications in which the saturation length is usually treated as an adjustable parameter \cite{Heetal09,Wuetal12}. The importance of the saturation length for such models has been debated among engineers \cite{DucRodi08,Caoetal11,Caoetal12}, and it was concluded that even at fluvial scale the influence of sediment transport saturation was significant \cite{Caoetal12}. We thus anticipate that the present work can provide substantial contributions to several areas of the geological, planetary and engineering sciences.

\
\section*{Acknowledgments}

We acknowledge support from Grants No. NSFC 41350110226, No. NSFC 41376095, No. ETH-10-09-2, No. NSF AGS 1137716, from the European Research Council (ERC) Advanced Grant No. 319968-FlowCCS, DFG through the Cluster of Excellence ''Engineering of Advanced Materials``, and the Collaborative Research Centre SFB814 (Additive Manufacturing). We further thank Miller Mendoza and Robert Sullivan for fruitful discussions and Orencio Dur\'an for allowing us to use his numerical model to confirm some of our arguments in Section~\ref{Derivation}.


\begin{thebibliography}{69}%
\makeatletter
\providecommand \@ifxundefined [1]{%
 \@ifx{#1\undefined}
}%
\providecommand \@ifnum [1]{%
 \ifnum #1\expandafter \@firstoftwo
 \else \expandafter \@secondoftwo
 \fi
}%
\providecommand \@ifx [1]{%
 \ifx #1\expandafter \@firstoftwo
 \else \expandafter \@secondoftwo
 \fi
}%
\providecommand \natexlab [1]{#1}%
\providecommand \enquote  [1]{``#1''}%
\providecommand \bibnamefont  [1]{#1}%
\providecommand \bibfnamefont [1]{#1}%
\providecommand \citenamefont [1]{#1}%
\providecommand \href@noop [0]{\@secondoftwo}%
\providecommand \href [0]{\begingroup \@sanitize@url \@href}%
\providecommand \@href[1]{\@@startlink{#1}\@@href}%
\providecommand \@@href[1]{\endgroup#1\@@endlink}%
\providecommand \@sanitize@url [0]{\catcode `\\12\catcode `\$12\catcode
  `\&12\catcode `\#12\catcode `\^12\catcode `\_12\catcode `\%12\relax}%
\providecommand \@@startlink[1]{}%
\providecommand \@@endlink[0]{}%
\providecommand \url  [0]{\begingroup\@sanitize@url \@url }%
\providecommand \@url [1]{\endgroup\@href {#1}{\urlprefix }}%
\providecommand \urlprefix  [0]{URL }%
\providecommand \Eprint [0]{\href }%
\providecommand \doibase [0]{http://dx.doi.org/}%
\providecommand \selectlanguage [0]{\@gobble}%
\providecommand \bibinfo  [0]{\@secondoftwo}%
\providecommand \bibfield  [0]{\@secondoftwo}%
\providecommand \translation [1]{[#1]}%
\providecommand \BibitemOpen [0]{}%
\providecommand \bibitemStop [0]{}%
\providecommand \bibitemNoStop [0]{.\EOS\space}%
\providecommand \EOS [0]{\spacefactor3000\relax}%
\providecommand \BibitemShut  [1]{\csname bibitem#1\endcsname}%
\let\auto@bib@innerbib\@empty
\bibitem [{\citenamefont {Bagnold}(1941)}]{Bagnold41}%
  \BibitemOpen
  \bibfield  {author} {\bibinfo {author} {\bibfnamefont {R.~A.}\ \bibnamefont
  {Bagnold}},\ }\href@noop {} {\emph {\bibinfo {title} {The physics of blown
  sand and desert dunes}}}\ (\bibinfo  {publisher} {Methuen, New York},\
  \bibinfo {year} {1941})\BibitemShut {NoStop}%
\bibitem [{\citenamefont {Rijn}(1993)}]{VanRijn93}%
  \BibitemOpen
  \bibfield  {author} {\bibinfo {author} {\bibfnamefont {L.~C.~V.}\
  \bibnamefont {Rijn}},\ }\href@noop {} {\emph {\bibinfo {title} {Principles of
  sediment transport in rivers, estuaries and coastal seas}}}\ (\bibinfo
  {publisher} {Aqua Publications, Amsterdam},\ \bibinfo {year}
  {1993})\BibitemShut {NoStop}%
\bibitem [{\citenamefont {Greeley}\ and\ \citenamefont
  {Iversen}(1985)}]{GreeleyIversen85}%
  \BibitemOpen
  \bibfield  {author} {\bibinfo {author} {\bibfnamefont {R.}~\bibnamefont
  {Greeley}}\ and\ \bibinfo {author} {\bibfnamefont {J.~D.}\ \bibnamefont
  {Iversen}},\ }\href@noop {} {\emph {\bibinfo {title} {Wind as a geological
  process on Earth, Mars, Venus, and Titan}}}\ (\bibinfo  {publisher}
  {Cambridge University Press},\ \bibinfo {year} {1985})\BibitemShut {NoStop}%
\bibitem [{\citenamefont {Shao}(2008)}]{Shao08}%
  \BibitemOpen
  \bibfield  {author} {\bibinfo {author} {\bibfnamefont {Y.}~\bibnamefont
  {Shao}},\ }\href@noop {} {\emph {\bibinfo {title} {Physics and modelling of
  wind erosion}}}\ (\bibinfo  {publisher} {Kluwer Academy, Dordrecht,
  Amsterdam},\ \bibinfo {year} {2008})\BibitemShut {NoStop}%
\bibitem [{\citenamefont {Kroy}\ \emph {et~al.}(2002)\citenamefont {Kroy},
  \citenamefont {Sauermann},\ and\ \citenamefont {Herrmann}}]{Kroyetal02a}%
  \BibitemOpen
  \bibfield  {author} {\bibinfo {author} {\bibfnamefont {K.}~\bibnamefont
  {Kroy}}, \bibinfo {author} {\bibfnamefont {G.}~\bibnamefont {Sauermann}}, \
  and\ \bibinfo {author} {\bibfnamefont {H.~J.}\ \bibnamefont {Herrmann}},\
  }\href {\doibase 10.1103/PhysRevE.66.031302} {\bibfield  {journal} {\bibinfo
  {journal} {Physical Review E}\ }\textbf {\bibinfo {volume} {66}},\ \bibinfo
  {pages} {031302} (\bibinfo {year} {2002})}\BibitemShut {NoStop}%
\bibitem [{\citenamefont {Kok}\ \emph {et~al.}(2012)\citenamefont {Kok},
  \citenamefont {Parteli}, \citenamefont {Michaels},\ and\ \citenamefont
  {Karam}}]{Koketal12}%
  \BibitemOpen
  \bibfield  {author} {\bibinfo {author} {\bibfnamefont {J.~F.}\ \bibnamefont
  {Kok}}, \bibinfo {author} {\bibfnamefont {E.~J.~R.}\ \bibnamefont {Parteli}},
  \bibinfo {author} {\bibfnamefont {T.~I.}\ \bibnamefont {Michaels}}, \ and\
  \bibinfo {author} {\bibfnamefont {D.~B.}\ \bibnamefont {Karam}},\ }\href@noop
  {} {\bibfield  {journal} {\bibinfo  {journal} {Reports on Progress in
  Physics}\ }\textbf {\bibinfo {volume} {75}},\ \bibinfo {pages} {106901}
  (\bibinfo {year} {2012})}\BibitemShut {NoStop}%
\bibitem [{\citenamefont {Bourke}\ \emph {et~al.}(2010)\citenamefont {Bourke},
  \citenamefont {Lancaster}, \citenamefont {Fenton}, \citenamefont {Parteli},
  \citenamefont {Zimbelman},\ and\ \citenamefont {Radebaugh}}]{Bourkeetal10}%
  \BibitemOpen
  \bibfield  {author} {\bibinfo {author} {\bibfnamefont {M.~C.}\ \bibnamefont
  {Bourke}}, \bibinfo {author} {\bibfnamefont {N.}~\bibnamefont {Lancaster}},
  \bibinfo {author} {\bibfnamefont {L.~K.}\ \bibnamefont {Fenton}}, \bibinfo
  {author} {\bibfnamefont {E.~J.~R.}\ \bibnamefont {Parteli}}, \bibinfo
  {author} {\bibfnamefont {J.~R.}\ \bibnamefont {Zimbelman}}, \ and\ \bibinfo
  {author} {\bibfnamefont {J.}~\bibnamefont {Radebaugh}},\ }\href@noop {}
  {\bibfield  {journal} {\bibinfo  {journal} {Geomorphology}\ }\textbf
  {\bibinfo {volume} {121}},\ \bibinfo {pages} {1} (\bibinfo {year}
  {2010})}\BibitemShut {NoStop}%
\bibitem [{\citenamefont {Ungar}\ and\ \citenamefont
  {Haff}(1987)}]{UngarHaff87}%
  \BibitemOpen
  \bibfield  {author} {\bibinfo {author} {\bibfnamefont {J.~E.}\ \bibnamefont
  {Ungar}}\ and\ \bibinfo {author} {\bibfnamefont {P.~K.}\ \bibnamefont
  {Haff}},\ }\href@noop {} {\bibfield  {journal} {\bibinfo  {journal}
  {Sedimentology}\ }\textbf {\bibinfo {volume} {34}},\ \bibinfo {pages} {289}
  (\bibinfo {year} {1987})}\BibitemShut {NoStop}%
\bibitem [{\citenamefont {Almeida}\ \emph {et~al.}(2007)\citenamefont
  {Almeida}, \citenamefont {Andrade},\ and\ \citenamefont
  {Herrmann}}]{Almeidaetal07}%
  \BibitemOpen
  \bibfield  {author} {\bibinfo {author} {\bibfnamefont {M.~P.}\ \bibnamefont
  {Almeida}}, \bibinfo {author} {\bibfnamefont {J.~S.}\ \bibnamefont
  {Andrade}}, \ and\ \bibinfo {author} {\bibfnamefont {H.~J.}\ \bibnamefont
  {Herrmann}},\ }\href@noop {} {\bibfield  {journal} {\bibinfo  {journal} {The
  European Physical Journal E}\ }\textbf {\bibinfo {volume} {22}},\ \bibinfo
  {pages} {195} (\bibinfo {year} {2007})}\BibitemShut {NoStop}%
\bibitem [{\citenamefont {Almeida}\ \emph {et~al.}(2008)\citenamefont
  {Almeida}, \citenamefont {Parteli}, \citenamefont {Andrade},\ and\
  \citenamefont {Herrmann}}]{Almeidaetal08}%
  \BibitemOpen
  \bibfield  {author} {\bibinfo {author} {\bibfnamefont {M.~P.}\ \bibnamefont
  {Almeida}}, \bibinfo {author} {\bibfnamefont {E.~J.~R.}\ \bibnamefont
  {Parteli}}, \bibinfo {author} {\bibfnamefont {J.~S.}\ \bibnamefont
  {Andrade}}, \ and\ \bibinfo {author} {\bibfnamefont {H.~J.}\ \bibnamefont
  {Herrmann}},\ }\href {\doibase 10.1073/pnas.0800202105} {\bibfield  {journal}
  {\bibinfo  {journal} {Proceedings of the National Academy of Science}\
  }\textbf {\bibinfo {volume} {105}},\ \bibinfo {pages} {6222} (\bibinfo {year}
  {2008})}\BibitemShut {NoStop}%
\bibitem [{\citenamefont {P\"ahtz}\ \emph {et~al.}(2012)\citenamefont
  {P\"ahtz}, \citenamefont {Kok},\ and\ \citenamefont
  {Herrmann}}]{Paehtzetal12a}%
  \BibitemOpen
  \bibfield  {author} {\bibinfo {author} {\bibfnamefont {T.}~\bibnamefont
  {P\"ahtz}}, \bibinfo {author} {\bibfnamefont {J.~F.}\ \bibnamefont {Kok}}, \
  and\ \bibinfo {author} {\bibfnamefont {H.~J.}\ \bibnamefont {Herrmann}},\
  }\href {\doibase 10.1088/1367-2630/14/4/043035} {\bibfield  {journal}
  {\bibinfo  {journal} {New Journal of Physics}\ }\textbf {\bibinfo {volume}
  {14}},\ \bibinfo {pages} {043035} (\bibinfo {year} {2012})}\BibitemShut
  {NoStop}%
\bibitem [{\citenamefont {Meyer-Peter}\ and\ \citenamefont
  {M\"uller}(1948)}]{MeyerPeterMuller48}%
  \BibitemOpen
  \bibfield  {author} {\bibinfo {author} {\bibfnamefont {E.}~\bibnamefont
  {Meyer-Peter}}\ and\ \bibinfo {author} {\bibfnamefont {R.}~\bibnamefont
  {M\"uller}},\ }in\ \href@noop {} {\emph {\bibinfo {booktitle} {Proceedings of
  the 2nd Meeting of the International Association for Hydraulic Structures
  Research}}}\ (\bibinfo {organization} {IAHR},\ \bibinfo {address}
  {Stockholm},\ \bibinfo {year} {1948})\BibitemShut {NoStop}%
\bibitem [{\citenamefont {Einstein}(1950)}]{Einstein50}%
  \BibitemOpen
  \bibfield  {author} {\bibinfo {author} {\bibfnamefont {H.~A.}\ \bibnamefont
  {Einstein}},\ }\href@noop {} {\emph {\bibinfo {title} {The bed-load function
  for sediment transportation in open channel flows}}}\ (\bibinfo  {publisher}
  {United States Department of Agriculture, Washington},\ \bibinfo {year}
  {1950})\BibitemShut {NoStop}%
\bibitem [{\citenamefont {Bagnold}(1966)}]{Bagnold66}%
  \BibitemOpen
  \bibfield  {author} {\bibinfo {author} {\bibfnamefont {R.~A.}\ \bibnamefont
  {Bagnold}},\ }in\ \href@noop {} {\emph {\bibinfo {booktitle} {US Geological
  Survey Professional Paper 422-I}}}\ (\bibinfo {year} {1966})\BibitemShut
  {NoStop}%
\bibitem [{\citenamefont {S{\o}rensen}(1991)}]{Sorensen91}%
  \BibitemOpen
  \bibfield  {author} {\bibinfo {author} {\bibfnamefont {M.}~\bibnamefont
  {S{\o}rensen}},\ }\href@noop {} {\bibfield  {journal} {\bibinfo  {journal}
  {Acta Mechanica Supplement}\ }\textbf {\bibinfo {volume} {1}},\ \bibinfo
  {pages} {67} (\bibinfo {year} {1991})}\BibitemShut {NoStop}%
\bibitem [{\citenamefont {Abrahams}\ and\ \citenamefont
  {Gao}(2006)}]{AbrahamsGao06}%
  \BibitemOpen
  \bibfield  {author} {\bibinfo {author} {\bibfnamefont {A.~D.}\ \bibnamefont
  {Abrahams}}\ and\ \bibinfo {author} {\bibfnamefont {P.}~\bibnamefont {Gao}},\
  }\href {\doibase 10.1002/esp.1300} {\bibfield  {journal} {\bibinfo  {journal}
  {Earth Surface Processes and Landforms}\ }\textbf {\bibinfo {volume} {31}},\
  \bibinfo {pages} {910} (\bibinfo {year} {2006})}\BibitemShut {NoStop}%
\bibitem [{\citenamefont {L\"ammel}\ \emph {et~al.}(2012)\citenamefont
  {L\"ammel}, \citenamefont {Rings},\ and\ \citenamefont
  {Kroy}}]{Laemmeletal12}%
  \BibitemOpen
  \bibfield  {author} {\bibinfo {author} {\bibfnamefont {M.}~\bibnamefont
  {L\"ammel}}, \bibinfo {author} {\bibfnamefont {D.}~\bibnamefont {Rings}}, \
  and\ \bibinfo {author} {\bibfnamefont {K.}~\bibnamefont {Kroy}},\ }\href@noop
  {} {\bibfield  {journal} {\bibinfo  {journal} {New Journal of Physics}\
  }\textbf {\bibinfo {volume} {14}},\ \bibinfo {pages} {093037} (\bibinfo
  {year} {2012})}\BibitemShut {NoStop}%
\bibitem [{\citenamefont {Sauermann}\ \emph {et~al.}(2001)\citenamefont
  {Sauermann}, \citenamefont {Kroy},\ and\ \citenamefont
  {Herrmann}}]{Sauermannetal01}%
  \BibitemOpen
  \bibfield  {author} {\bibinfo {author} {\bibfnamefont {G.}~\bibnamefont
  {Sauermann}}, \bibinfo {author} {\bibfnamefont {K.}~\bibnamefont {Kroy}}, \
  and\ \bibinfo {author} {\bibfnamefont {H.~J.}\ \bibnamefont {Herrmann}},\
  }\href {\doibase 10.1103/PhysRevE.64.031305} {\bibfield  {journal} {\bibinfo
  {journal} {Physical Review E}\ }\textbf {\bibinfo {volume} {64}},\ \bibinfo
  {pages} {31305} (\bibinfo {year} {2001})}\BibitemShut {NoStop}%
\bibitem [{\citenamefont {Claudin}\ and\ \citenamefont
  {Andreotti}(2006)}]{ClaudinAndreotti06}%
  \BibitemOpen
  \bibfield  {author} {\bibinfo {author} {\bibfnamefont {P.}~\bibnamefont
  {Claudin}}\ and\ \bibinfo {author} {\bibfnamefont {B.}~\bibnamefont
  {Andreotti}},\ }\href {\doibase 10.1016/j.epsl.2006.09.004} {\bibfield
  {journal} {\bibinfo  {journal} {Earth and Planetary Science Letters}\
  }\textbf {\bibinfo {volume} {252}},\ \bibinfo {pages} {30} (\bibinfo {year}
  {2006})}\BibitemShut {NoStop}%
\bibitem [{\citenamefont {Andreotti}\ \emph {et~al.}(2010)\citenamefont
  {Andreotti}, \citenamefont {Claudin},\ and\ \citenamefont
  {Pouliquen}}]{Andreottietal10}%
  \BibitemOpen
  \bibfield  {author} {\bibinfo {author} {\bibfnamefont {B.}~\bibnamefont
  {Andreotti}}, \bibinfo {author} {\bibfnamefont {P.}~\bibnamefont {Claudin}},
  \ and\ \bibinfo {author} {\bibfnamefont {O.}~\bibnamefont {Pouliquen}},\
  }\href@noop {} {\bibfield  {journal} {\bibinfo  {journal} {Geomorphology}\
  }\textbf {\bibinfo {volume} {123}},\ \bibinfo {pages} {343} (\bibinfo {year}
  {2010})}\BibitemShut {NoStop}%
\bibitem [{\citenamefont {Fourri\`ere}\ \emph {et~al.}(2010)\citenamefont
  {Fourri\`ere}, \citenamefont {Claudin},\ and\ \citenamefont
  {Andreotti}}]{Fourriereetal10}%
  \BibitemOpen
  \bibfield  {author} {\bibinfo {author} {\bibfnamefont {A.}~\bibnamefont
  {Fourri\`ere}}, \bibinfo {author} {\bibfnamefont {P.}~\bibnamefont
  {Claudin}}, \ and\ \bibinfo {author} {\bibfnamefont {B.}~\bibnamefont
  {Andreotti}},\ }\href@noop {} {\bibfield  {journal} {\bibinfo  {journal}
  {Journal of Fluid Mechanics}\ }\textbf {\bibinfo {volume} {649}},\ \bibinfo
  {pages} {287} (\bibinfo {year} {2010})}\BibitemShut {NoStop}%
\bibitem [{\citenamefont {Hersen}\ \emph {et~al.}(2002)\citenamefont {Hersen},
  \citenamefont {Douady},\ and\ \citenamefont {Andreotti}}]{Hersenetal02}%
  \BibitemOpen
  \bibfield  {author} {\bibinfo {author} {\bibfnamefont {P.}~\bibnamefont
  {Hersen}}, \bibinfo {author} {\bibfnamefont {S.}~\bibnamefont {Douady}}, \
  and\ \bibinfo {author} {\bibfnamefont {B.}~\bibnamefont {Andreotti}},\
  }\href@noop {} {\bibfield  {journal} {\bibinfo  {journal} {Physical Review
  Letters}\ }\textbf {\bibinfo {volume} {89}},\ \bibinfo {pages} {264301}
  (\bibinfo {year} {2002})}\BibitemShut {NoStop}%
\bibitem [{\citenamefont {Franklin}\ and\ \citenamefont
  {Charru}(2011)}]{FranklinCharru11}%
  \BibitemOpen
  \bibfield  {author} {\bibinfo {author} {\bibfnamefont {E.~M.}\ \bibnamefont
  {Franklin}}\ and\ \bibinfo {author} {\bibfnamefont {F.}~\bibnamefont
  {Charru}},\ }\href@noop {} {\bibfield  {journal} {\bibinfo  {journal}
  {Journal of Fluid Mechanics}\ }\textbf {\bibinfo {volume} {675}},\ \bibinfo
  {pages} {199} (\bibinfo {year} {2011})}\BibitemShut {NoStop}%
\bibitem [{\citenamefont {Charru}(2006)}]{Charru06}%
  \BibitemOpen
  \bibfield  {author} {\bibinfo {author} {\bibfnamefont {F.}~\bibnamefont
  {Charru}},\ }\href {\doibase 10.1063/1.2397005} {\bibfield  {journal}
  {\bibinfo  {journal} {Physics of Fluids}\ }\textbf {\bibinfo {volume} {18}},\
  \bibinfo {pages} {121508} (\bibinfo {year} {2006})}\BibitemShut {NoStop}%
\bibitem [{\citenamefont {Ma}\ and\ \citenamefont {Zheng}(2011)}]{MaZheng11}%
  \BibitemOpen
  \bibfield  {author} {\bibinfo {author} {\bibfnamefont {G.~S.}\ \bibnamefont
  {Ma}}\ and\ \bibinfo {author} {\bibfnamefont {X.~J.}\ \bibnamefont {Zheng}},\
  }\href@noop {} {\bibfield  {journal} {\bibinfo  {journal} {The European
  Physical Journal E}\ }\textbf {\bibinfo {volume} {34}},\ \bibinfo {pages} {1}
  (\bibinfo {year} {2011})}\BibitemShut {NoStop}%
\bibitem [{\citenamefont {P\"ahtz}\ \emph {et~al.}(2013)\citenamefont
  {P\"ahtz}, \citenamefont {Kok}, \citenamefont {Parteli},\ and\ \citenamefont
  {Herrmann}}]{Paehtzetal13}%
  \BibitemOpen
  \bibfield  {author} {\bibinfo {author} {\bibfnamefont {T.}~\bibnamefont
  {P\"ahtz}}, \bibinfo {author} {\bibfnamefont {J.~F.}\ \bibnamefont {Kok}},
  \bibinfo {author} {\bibfnamefont {E.~J.~R.}\ \bibnamefont {Parteli}}, \ and\
  \bibinfo {author} {\bibfnamefont {H.~J.}\ \bibnamefont {Herrmann}},\ }\href
  {\doibase 10.1103/PhysRevLett.111.218002} {\bibfield  {journal} {\bibinfo
  {journal} {Physical Review Letters}\ }\textbf {\bibinfo {volume} {111}},\
  \bibinfo {pages} {218002} (\bibinfo {year} {2013})}\BibitemShut {NoStop}%
\bibitem [{\citenamefont {Moraga}\ \emph {et~al.}(1999)\citenamefont {Moraga},
  \citenamefont {Bonetto},\ and\ \citenamefont {Lahey}}]{Moragaetal99}%
  \BibitemOpen
  \bibfield  {author} {\bibinfo {author} {\bibfnamefont {F.~J.}\ \bibnamefont
  {Moraga}}, \bibinfo {author} {\bibfnamefont {F.~J.}\ \bibnamefont {Bonetto}},
  \ and\ \bibinfo {author} {\bibfnamefont {R.~T.}\ \bibnamefont {Lahey}},\
  }\href@noop {} {\bibfield  {journal} {\bibinfo  {journal} {International
  Journal of Multiphase Flow}\ }\textbf {\bibinfo {volume} {25}},\ \bibinfo
  {pages} {1321} (\bibinfo {year} {1999})}\BibitemShut {NoStop}%
\bibitem [{\citenamefont {Nino}\ and\ \citenamefont
  {Garcia}(1998)}]{NinoGarcia98a}%
  \BibitemOpen
  \bibfield  {author} {\bibinfo {author} {\bibfnamefont {Y.}~\bibnamefont
  {Nino}}\ and\ \bibinfo {author} {\bibfnamefont {M.}~\bibnamefont {Garcia}},\
  }\href@noop {} {\bibfield  {journal} {\bibinfo  {journal} {Hydrological
  Processes}\ }\textbf {\bibinfo {volume} {12}},\ \bibinfo {pages} {1197}
  (\bibinfo {year} {1998})}\BibitemShut {NoStop}%
\bibitem [{\citenamefont {Gao}(2008)}]{Gao08}%
  \BibitemOpen
  \bibfield  {author} {\bibinfo {author} {\bibfnamefont {P.}~\bibnamefont
  {Gao}},\ }\href {\doibase 10.1061/(ASCE)0733-9429(2008)(134:3)(340)}
  {\bibfield  {journal} {\bibinfo  {journal} {Journal of Hydraulic
  Engineering}\ }\textbf {\bibinfo {volume} {134}},\ \bibinfo {pages} {340}
  (\bibinfo {year} {2008})}\BibitemShut {NoStop}%
\bibitem [{\citenamefont {Dur\'an}\ \emph {et~al.}(2012)\citenamefont
  {Dur\'an}, \citenamefont {Andreotti},\ and\ \citenamefont
  {Claudin}}]{Duranetal12}%
  \BibitemOpen
  \bibfield  {author} {\bibinfo {author} {\bibfnamefont {O.}~\bibnamefont
  {Dur\'an}}, \bibinfo {author} {\bibfnamefont {B.}~\bibnamefont {Andreotti}},
  \ and\ \bibinfo {author} {\bibfnamefont {P.}~\bibnamefont {Claudin}},\
  }\href@noop {} {\bibfield  {journal} {\bibinfo  {journal} {Physics of
  Fluids}\ }\textbf {\bibinfo {volume} {24}},\ \bibinfo {pages} {103306}
  (\bibinfo {year} {2012})}\BibitemShut {NoStop}%
\bibitem [{\citenamefont {Babic}(1997)}]{Babic97}%
  \BibitemOpen
  \bibfield  {author} {\bibinfo {author} {\bibfnamefont {M.}~\bibnamefont
  {Babic}},\ }\href@noop {} {\bibfield  {journal} {\bibinfo  {journal}
  {International Journal of Engineering Science}\ }\textbf {\bibinfo {volume}
  {35}},\ \bibinfo {pages} {523} (\bibinfo {year} {1997})}\BibitemShut
  {NoStop}%
\bibitem [{\citenamefont {Bagnold}(1956)}]{Bagnold56}%
  \BibitemOpen
  \bibfield  {author} {\bibinfo {author} {\bibfnamefont {R.~A.}\ \bibnamefont
  {Bagnold}},\ }\href@noop {} {\bibfield  {journal} {\bibinfo  {journal}
  {Philosophical Transactions of the Royal Society London A}\ }\textbf
  {\bibinfo {volume} {249}},\ \bibinfo {pages} {235} (\bibinfo {year}
  {1956})}\BibitemShut {NoStop}%
\bibitem [{\citenamefont {Bagnold}(1973)}]{Bagnold73}%
  \BibitemOpen
  \bibfield  {author} {\bibinfo {author} {\bibfnamefont {R.~A.}\ \bibnamefont
  {Bagnold}},\ }\href@noop {} {\bibfield  {journal} {\bibinfo  {journal}
  {Proceedings of the Royal Society London Series A}\ }\textbf {\bibinfo
  {volume} {332}},\ \bibinfo {pages} {473} (\bibinfo {year}
  {1973})}\BibitemShut {NoStop}%
\bibitem [{\citenamefont {Ashida}\ and\ \citenamefont
  {Michiue}(1972)}]{AshidaMichiue72}%
  \BibitemOpen
  \bibfield  {author} {\bibinfo {author} {\bibfnamefont {K.}~\bibnamefont
  {Ashida}}\ and\ \bibinfo {author} {\bibfnamefont {M.}~\bibnamefont
  {Michiue}},\ }in\ \href@noop {} {\emph {\bibinfo {booktitle} {Transcripts of
  the Japan Society for Civil Engineers}}},\ Vol.\ \bibinfo {volume} {206}\
  (\bibinfo {year} {1972})\ pp.\ \bibinfo {pages} {59--69}\BibitemShut
  {NoStop}%
\bibitem [{\citenamefont {Zhang}\ and\ \citenamefont
  {Campbell}(1992)}]{ZhangCampbell92}%
  \BibitemOpen
  \bibfield  {author} {\bibinfo {author} {\bibfnamefont {Y.}~\bibnamefont
  {Zhang}}\ and\ \bibinfo {author} {\bibfnamefont {C.~S.}\ \bibnamefont
  {Campbell}},\ }\href@noop {} {\bibfield  {journal} {\bibinfo  {journal}
  {Journal of Fluid Mechanics}\ }\textbf {\bibinfo {volume} {237}},\ \bibinfo
  {pages} {541} (\bibinfo {year} {1992})}\BibitemShut {NoStop}%
\bibitem [{\citenamefont {Du}\ \emph {et~al.}(2006)\citenamefont {Du},
  \citenamefont {Bao}, \citenamefont {Xu},\ and\ \citenamefont
  {Wei}}]{Duetal06}%
  \BibitemOpen
  \bibfield  {author} {\bibinfo {author} {\bibfnamefont {W.}~\bibnamefont
  {Du}}, \bibinfo {author} {\bibfnamefont {X.}~\bibnamefont {Bao}}, \bibinfo
  {author} {\bibfnamefont {J.}~\bibnamefont {Xu}}, \ and\ \bibinfo {author}
  {\bibfnamefont {W.}~\bibnamefont {Wei}},\ }\href@noop {} {\bibfield
  {journal} {\bibinfo  {journal} {Chemical Engineering Science}\ }\textbf
  {\bibinfo {volume} {61}},\ \bibinfo {pages} {1401} (\bibinfo {year}
  {2006})}\BibitemShut {NoStop}%
\bibitem [{\citenamefont {Julien}(1995)}]{Julien95}%
  \BibitemOpen
  \bibfield  {author} {\bibinfo {author} {\bibfnamefont {P.~Y.}\ \bibnamefont
  {Julien}},\ }\href@noop {} {\emph {\bibinfo {title} {Erosion and
  Sedimentation}}}\ (\bibinfo  {publisher} {Press Syndicate of the University
  of Cambridge},\ \bibinfo {year} {1995})\BibitemShut {NoStop}%
\bibitem [{\citenamefont {George}(2009)}]{George09}%
  \BibitemOpen
  \bibfield  {author} {\bibinfo {author} {\bibfnamefont {W.~K.}\ \bibnamefont
  {George}},\ }\href@noop {} {\emph {\bibinfo {title} {Lectures in turbulence
  for the 21st Century}}}\ (\bibinfo  {publisher} {Chalmers University
  Gothenborg},\ \bibinfo {year} {2009})\BibitemShut {NoStop}%
\bibitem [{\citenamefont {Nino}\ and\ \citenamefont
  {Garcia}(1994)}]{NinoGarcia94}%
  \BibitemOpen
  \bibfield  {author} {\bibinfo {author} {\bibfnamefont {Y.}~\bibnamefont
  {Nino}}\ and\ \bibinfo {author} {\bibfnamefont {M.}~\bibnamefont {Garcia}},\
  }\href@noop {} {\bibfield  {journal} {\bibinfo  {journal} {Water Resources
  Research}\ }\textbf {\bibinfo {volume} {30}},\ \bibinfo {pages} {1915}
  (\bibinfo {year} {1994})}\BibitemShut {NoStop}%
\bibitem [{\citenamefont {Seminara}\ \emph {et~al.}(2002)\citenamefont
  {Seminara}, \citenamefont {Solari},\ and\ \citenamefont
  {Parker}}]{Seminaraetal02}%
  \BibitemOpen
  \bibfield  {author} {\bibinfo {author} {\bibfnamefont {G.}~\bibnamefont
  {Seminara}}, \bibinfo {author} {\bibfnamefont {L.}~\bibnamefont {Solari}}, \
  and\ \bibinfo {author} {\bibfnamefont {G.}~\bibnamefont {Parker}},\ }\href
  {\doibase 10.1029/2001WR000681} {\bibfield  {journal} {\bibinfo  {journal}
  {Water Resources Research}\ }\textbf {\bibinfo {volume} {38}},\ \bibinfo
  {pages} {1249} (\bibinfo {year} {2002})}\BibitemShut {NoStop}%
\bibitem [{\citenamefont {Lajeunesse}\ \emph {et~al.}(2010)\citenamefont
  {Lajeunesse}, \citenamefont {Malverti},\ and\ \citenamefont
  {Charru}}]{Lajeunesseetal10}%
  \BibitemOpen
  \bibfield  {author} {\bibinfo {author} {\bibfnamefont {E.}~\bibnamefont
  {Lajeunesse}}, \bibinfo {author} {\bibfnamefont {L.}~\bibnamefont
  {Malverti}}, \ and\ \bibinfo {author} {\bibfnamefont {F.}~\bibnamefont
  {Charru}},\ }\href {\doibase 10.1029/2009JF001628} {\bibfield  {journal}
  {\bibinfo  {journal} {Journal of Geophysical Research}\ }\textbf {\bibinfo
  {volume} {115}},\ \bibinfo {pages} {F04001} (\bibinfo {year}
  {2010})}\BibitemShut {NoStop}%
\bibitem [{\citenamefont {Francis}(1973)}]{Francis73}%
  \BibitemOpen
  \bibfield  {author} {\bibinfo {author} {\bibfnamefont {J.~R.~D.}\
  \bibnamefont {Francis}},\ }\href@noop {} {\bibfield  {journal} {\bibinfo
  {journal} {Philosophical Transactions of the Royal Society London A}\
  }\textbf {\bibinfo {volume} {332}},\ \bibinfo {pages} {443} (\bibinfo {year}
  {1973})}\BibitemShut {NoStop}%
\bibitem [{\citenamefont {Abbott}\ and\ \citenamefont
  {Francis}(1977)}]{AbbottFrancis77}%
  \BibitemOpen
  \bibfield  {author} {\bibinfo {author} {\bibfnamefont {J.~E.}\ \bibnamefont
  {Abbott}}\ and\ \bibinfo {author} {\bibfnamefont {J.~R.~D.}\ \bibnamefont
  {Francis}},\ }\href {\doibase 10.1098/rsta.1977.0009} {\bibfield  {journal}
  {\bibinfo  {journal} {Philosophical Transactions of the Royal Society London
  A}\ }\textbf {\bibinfo {volume} {284}},\ \bibinfo {pages} {225} (\bibinfo
  {year} {1977})}\BibitemShut {NoStop}%
\bibitem [{\citenamefont {Nino}\ \emph {et~al.}(1994)\citenamefont {Nino},
  \citenamefont {Garcia},\ and\ \citenamefont {Ayala}}]{Ninoetal94}%
  \BibitemOpen
  \bibfield  {author} {\bibinfo {author} {\bibfnamefont {Y.}~\bibnamefont
  {Nino}}, \bibinfo {author} {\bibfnamefont {M.}~\bibnamefont {Garcia}}, \ and\
  \bibinfo {author} {\bibfnamefont {L.}~\bibnamefont {Ayala}},\ }\href@noop {}
  {\bibfield  {journal} {\bibinfo  {journal} {Water Resources Research}\
  }\textbf {\bibinfo {volume} {30}},\ \bibinfo {pages} {1907} (\bibinfo {year}
  {1994})}\BibitemShut {NoStop}%
\bibitem [{\citenamefont {Creyssels}\ \emph {et~al.}(2009)\citenamefont
  {Creyssels}, \citenamefont {Dupont}, \citenamefont {el~Moctar}, \citenamefont
  {Valance}, \citenamefont {Cantat}, \citenamefont {Jenkins}, \citenamefont
  {Pasini},\ and\ \citenamefont {Rasmussen}}]{Creysselsetal09}%
  \BibitemOpen
  \bibfield  {author} {\bibinfo {author} {\bibfnamefont {M.}~\bibnamefont
  {Creyssels}}, \bibinfo {author} {\bibfnamefont {P.}~\bibnamefont {Dupont}},
  \bibinfo {author} {\bibfnamefont {A.~O.}\ \bibnamefont {el~Moctar}}, \bibinfo
  {author} {\bibfnamefont {A.}~\bibnamefont {Valance}}, \bibinfo {author}
  {\bibfnamefont {I.}~\bibnamefont {Cantat}}, \bibinfo {author} {\bibfnamefont
  {J.~T.}\ \bibnamefont {Jenkins}}, \bibinfo {author} {\bibfnamefont {J.~M.}\
  \bibnamefont {Pasini}}, \ and\ \bibinfo {author} {\bibfnamefont {K.~R.}\
  \bibnamefont {Rasmussen}},\ }\href {\doibase 10.1017/S0022112008005491}
  {\bibfield  {journal} {\bibinfo  {journal} {Journal of Fluid Mechanics}\
  }\textbf {\bibinfo {volume} {625}},\ \bibinfo {pages} {47} (\bibinfo {year}
  {2009})}\BibitemShut {NoStop}%
\bibitem [{\citenamefont {Greeley}\ \emph {et~al.}(1996)\citenamefont
  {Greeley}, \citenamefont {Blumberg},\ and\ \citenamefont
  {Williams}}]{Greeleyetal96}%
  \BibitemOpen
  \bibfield  {author} {\bibinfo {author} {\bibfnamefont {R.}~\bibnamefont
  {Greeley}}, \bibinfo {author} {\bibfnamefont {D.~G.}\ \bibnamefont
  {Blumberg}}, \ and\ \bibinfo {author} {\bibfnamefont {S.~H.}\ \bibnamefont
  {Williams}},\ }\href@noop {} {\bibfield  {journal} {\bibinfo  {journal}
  {Sedimentology}\ }\textbf {\bibinfo {volume} {43}},\ \bibinfo {pages} {41}
  (\bibinfo {year} {1996})}\BibitemShut {NoStop}%
\bibitem [{\citenamefont {Rasmussen}\ and\ \citenamefont
  {S{\o}rensen}(2008)}]{RasmussenSorensen08}%
  \BibitemOpen
  \bibfield  {author} {\bibinfo {author} {\bibfnamefont {K.~R.}\ \bibnamefont
  {Rasmussen}}\ and\ \bibinfo {author} {\bibfnamefont {M.}~\bibnamefont
  {S{\o}rensen}},\ }\href {\doibase 10.1029/2007JF000774} {\bibfield  {journal}
  {\bibinfo  {journal} {Journal of Geophysical Research}\ }\textbf {\bibinfo
  {volume} {113}},\ \bibinfo {pages} {F02S12} (\bibinfo {year}
  {2008})}\BibitemShut {NoStop}%
\bibitem [{\citenamefont {Kok}\ and\ \citenamefont {Renno}(2009)}]{KokRenno09}%
  \BibitemOpen
  \bibfield  {author} {\bibinfo {author} {\bibfnamefont {J.~F.}\ \bibnamefont
  {Kok}}\ and\ \bibinfo {author} {\bibfnamefont {N.~O.}\ \bibnamefont
  {Renno}},\ }\href {\doibase 10.1029/2009JD011702} {\bibfield  {journal}
  {\bibinfo  {journal} {Journal of Geophysical Research}\ }\textbf {\bibinfo
  {volume} {114}},\ \bibinfo {pages} {D17204} (\bibinfo {year}
  {2009})}\BibitemShut {NoStop}%
\bibitem [{\citenamefont {Carneiro}\ \emph {et~al.}(2011)\citenamefont
  {Carneiro}, \citenamefont {P\"ahtz},\ and\ \citenamefont
  {Herrmann}}]{Carneiroetal11}%
  \BibitemOpen
  \bibfield  {author} {\bibinfo {author} {\bibfnamefont {M.~V.}\ \bibnamefont
  {Carneiro}}, \bibinfo {author} {\bibfnamefont {T.}~\bibnamefont {P\"ahtz}}, \
  and\ \bibinfo {author} {\bibfnamefont {H.~J.}\ \bibnamefont {Herrmann}},\
  }\href {\doibase 10.1103/PhysRevLett.107.098001} {\bibfield  {journal}
  {\bibinfo  {journal} {Physical Review Letters}\ }\textbf {\bibinfo {volume}
  {107}},\ \bibinfo {pages} {098001} (\bibinfo {year} {2011})}\BibitemShut
  {NoStop}%
\bibitem [{\citenamefont {Carneiro}\ \emph {et~al.}(2013)\citenamefont
  {Carneiro}, \citenamefont {Ara\'ujo}, \citenamefont {P\"ahtz},\ and\
  \citenamefont {Herrmann}}]{Carneiroetal13}%
  \BibitemOpen
  \bibfield  {author} {\bibinfo {author} {\bibfnamefont {M.~V.}\ \bibnamefont
  {Carneiro}}, \bibinfo {author} {\bibfnamefont {N.~A.~M.}\ \bibnamefont
  {Ara\'ujo}}, \bibinfo {author} {\bibfnamefont {T.}~\bibnamefont {P\"ahtz}}, \
  and\ \bibinfo {author} {\bibfnamefont {H.~J.}\ \bibnamefont {Herrmann}},\
  }\href {\doibase 10.1103/PhysRevLett.111.058001} {\bibfield  {journal}
  {\bibinfo  {journal} {Physical Review Letters}\ }\textbf {\bibinfo {volume}
  {111}},\ \bibinfo {pages} {058001} (\bibinfo {year} {2013})}\BibitemShut
  {NoStop}%
\bibitem [{\citenamefont {Andreotti}(2004)}]{Andreotti04}%
  \BibitemOpen
  \bibfield  {author} {\bibinfo {author} {\bibfnamefont {B.}~\bibnamefont
  {Andreotti}},\ }\href {\doibase 10.1017/S0022112004009073} {\bibfield
  {journal} {\bibinfo  {journal} {Journal of Fluid Mechanics}\ }\textbf
  {\bibinfo {volume} {510}},\ \bibinfo {pages} {47} (\bibinfo {year}
  {2004})}\BibitemShut {NoStop}%
\bibitem [{\citenamefont {Beladjine}\ \emph {et~al.}(2007)\citenamefont
  {Beladjine}, \citenamefont {Ammi}, \citenamefont {Oger},\ and\ \citenamefont
  {Valance}}]{Beladjineetal07}%
  \BibitemOpen
  \bibfield  {author} {\bibinfo {author} {\bibfnamefont {D.}~\bibnamefont
  {Beladjine}}, \bibinfo {author} {\bibfnamefont {M.}~\bibnamefont {Ammi}},
  \bibinfo {author} {\bibfnamefont {L.}~\bibnamefont {Oger}}, \ and\ \bibinfo
  {author} {\bibfnamefont {A.}~\bibnamefont {Valance}},\ }\href@noop {}
  {\bibfield  {journal} {\bibinfo  {journal} {Physical Review E}\ }\textbf
  {\bibinfo {volume} {75}},\ \bibinfo {pages} {061305} (\bibinfo {year}
  {2007})}\BibitemShut {NoStop}%
\bibitem [{\citenamefont {Oger}\ \emph {et~al.}(2008)\citenamefont {Oger},
  \citenamefont {Ammi}, \citenamefont {Valance},\ and\ \citenamefont
  {Beladjine}}]{Ogeretal08}%
  \BibitemOpen
  \bibfield  {author} {\bibinfo {author} {\bibfnamefont {L.}~\bibnamefont
  {Oger}}, \bibinfo {author} {\bibfnamefont {M.}~\bibnamefont {Ammi}}, \bibinfo
  {author} {\bibfnamefont {A.}~\bibnamefont {Valance}}, \ and\ \bibinfo
  {author} {\bibfnamefont {D.}~\bibnamefont {Beladjine}},\ }\href@noop {}
  {\bibfield  {journal} {\bibinfo  {journal} {Computers and Mathematics with
  Applications}\ }\textbf {\bibinfo {volume} {55}},\ \bibinfo {pages} {132}
  (\bibinfo {year} {2008})}\BibitemShut {NoStop}%
\bibitem [{\citenamefont {Cheng}\ and\ \citenamefont
  {Chiew}(1998)}]{ChengChiew98}%
  \BibitemOpen
  \bibfield  {author} {\bibinfo {author} {\bibfnamefont {N.~S.}\ \bibnamefont
  {Cheng}}\ and\ \bibinfo {author} {\bibfnamefont {Y.~M.}\ \bibnamefont
  {Chiew}},\ }\href {\doibase 10.1061/(ASCE)0733-9429(1998)124:12(1235)}
  {\bibfield  {journal} {\bibinfo  {journal} {Journal of Hydraulic
  Engineering}\ }\textbf {\bibinfo {volume} {124}},\ \bibinfo {pages} {1235}
  (\bibinfo {year} {1998})}\BibitemShut {NoStop}%
\bibitem [{\citenamefont {Kok}(2010)}]{Kok10a}%
  \BibitemOpen
  \bibfield  {author} {\bibinfo {author} {\bibfnamefont {J.~F.}\ \bibnamefont
  {Kok}},\ }\href {\doibase 10.1103/PhysRevLett.104.074502} {\bibfield
  {journal} {\bibinfo  {journal} {Physical Review Letters}\ }\textbf {\bibinfo
  {volume} {104}},\ \bibinfo {pages} {074502} (\bibinfo {year}
  {2010})}\BibitemShut {NoStop}%
\bibitem [{\citenamefont {Luque}\ and\ \citenamefont {van
  Beek}(1976)}]{FernandezLuqueBeek76}%
  \BibitemOpen
  \bibfield  {author} {\bibinfo {author} {\bibfnamefont {R.~F.}\ \bibnamefont
  {Luque}}\ and\ \bibinfo {author} {\bibfnamefont {R.}~\bibnamefont {van
  Beek}},\ }\href@noop {} {\bibfield  {journal} {\bibinfo  {journal} {Journal
  of Hydraulic Research}\ }\textbf {\bibinfo {volume} {14}},\ \bibinfo {pages}
  {127} (\bibinfo {year} {1976})}\BibitemShut {NoStop}%
\bibitem [{\citenamefont {Rijn}(1984)}]{VanRijn84}%
  \BibitemOpen
  \bibfield  {author} {\bibinfo {author} {\bibfnamefont {L.~C.~V.}\
  \bibnamefont {Rijn}},\ }\href@noop {} {\bibfield  {journal} {\bibinfo
  {journal} {Journal of Hydraulic Engineering}\ }\textbf {\bibinfo {volume}
  {110}},\ \bibinfo {pages} {1431} (\bibinfo {year} {1984})}\BibitemShut
  {NoStop}%
\bibitem [{\citenamefont {Hu}\ and\ \citenamefont {Hui}(1996)}]{HuHui96}%
  \BibitemOpen
  \bibfield  {author} {\bibinfo {author} {\bibfnamefont {C.}~\bibnamefont
  {Hu}}\ and\ \bibinfo {author} {\bibfnamefont {Y.}~\bibnamefont {Hui}},\
  }\href@noop {} {\bibfield  {journal} {\bibinfo  {journal} {Journal of
  Hydraulic Engineering}\ }\textbf {\bibinfo {volume} {122}},\ \bibinfo {pages}
  {245} (\bibinfo {year} {1996})}\BibitemShut {NoStop}%
\bibitem [{\citenamefont {Paphitis}(2001)}]{Paphitis01}%
  \BibitemOpen
  \bibfield  {author} {\bibinfo {author} {\bibfnamefont {D.}~\bibnamefont
  {Paphitis}},\ }\href@noop {} {\bibfield  {journal} {\bibinfo  {journal}
  {Coastal Engineering}\ }\textbf {\bibinfo {volume} {43}},\ \bibinfo {pages}
  {227} (\bibinfo {year} {2001})}\BibitemShut {NoStop}%
\bibitem [{\citenamefont {Parteli}\ \emph {et~al.}(2007)\citenamefont
  {Parteli}, \citenamefont {Dur\'an},\ and\ \citenamefont
  {Herrmann}}]{Partelietal07}%
  \BibitemOpen
  \bibfield  {author} {\bibinfo {author} {\bibfnamefont {E.~J.~R.}\
  \bibnamefont {Parteli}}, \bibinfo {author} {\bibfnamefont {O.}~\bibnamefont
  {Dur\'an}}, \ and\ \bibinfo {author} {\bibfnamefont {H.~J.}\ \bibnamefont
  {Herrmann}},\ }\href {\doibase 10.1103/PhysRevE.75.011301} {\bibfield
  {journal} {\bibinfo  {journal} {Physical Review E}\ }\textbf {\bibinfo
  {volume} {75}},\ \bibinfo {pages} {011301} (\bibinfo {year}
  {2007})}\BibitemShut {NoStop}%
\bibitem [{\citenamefont {Parteli}\ and\ \citenamefont
  {Herrmann}(2007)}]{ParteliHerrmann07b}%
  \BibitemOpen
  \bibfield  {author} {\bibinfo {author} {\bibfnamefont {E.~J.~R.}\
  \bibnamefont {Parteli}}\ and\ \bibinfo {author} {\bibfnamefont {H.~J.}\
  \bibnamefont {Herrmann}},\ }\href {\doibase 10.1103/PhysRevLett.98.198001}
  {\bibfield  {journal} {\bibinfo  {journal} {Physical Review Letters}\
  }\textbf {\bibinfo {volume} {98}},\ \bibinfo {pages} {198001} (\bibinfo
  {year} {2007})}\BibitemShut {NoStop}%
\bibitem [{\citenamefont {Dur\'an}\ \emph {et~al.}(2011)\citenamefont
  {Dur\'an}, \citenamefont {Claudin},\ and\ \citenamefont
  {Andreotti}}]{Duranetal11}%
  \BibitemOpen
  \bibfield  {author} {\bibinfo {author} {\bibfnamefont {O.}~\bibnamefont
  {Dur\'an}}, \bibinfo {author} {\bibfnamefont {P.}~\bibnamefont {Claudin}}, \
  and\ \bibinfo {author} {\bibfnamefont {B.}~\bibnamefont {Andreotti}},\ }\href
  {\doibase 10.1016/j.aeolia.2011.07.006} {\bibfield  {journal} {\bibinfo
  {journal} {Aeolian Research}\ }\textbf {\bibinfo {volume} {3}},\ \bibinfo
  {pages} {243} (\bibinfo {year} {2011})}\BibitemShut {NoStop}%
\bibitem [{\citenamefont {Jackson}\ and\ \citenamefont
  {Hunt}(1975)}]{JacksonHunt75}%
  \BibitemOpen
  \bibfield  {author} {\bibinfo {author} {\bibfnamefont {P.~S.}\ \bibnamefont
  {Jackson}}\ and\ \bibinfo {author} {\bibfnamefont {J.~C.~R.}\ \bibnamefont
  {Hunt}},\ }\href@noop {} {\bibfield  {journal} {\bibinfo  {journal}
  {Quarterly Journal of the Royal Meteorological Society}\ }\textbf {\bibinfo
  {volume} {101}},\ \bibinfo {pages} {929} (\bibinfo {year}
  {1975})}\BibitemShut {NoStop}%
\bibitem [{\citenamefont {Hunt}\ \emph {et~al.}(1988)\citenamefont {Hunt},
  \citenamefont {Leibovich},\ and\ \citenamefont {Richards}}]{Huntetal88}%
  \BibitemOpen
  \bibfield  {author} {\bibinfo {author} {\bibfnamefont {J.~C.~R.}\
  \bibnamefont {Hunt}}, \bibinfo {author} {\bibfnamefont {S.}~\bibnamefont
  {Leibovich}}, \ and\ \bibinfo {author} {\bibfnamefont {K.~J.}\ \bibnamefont
  {Richards}},\ }\href@noop {} {\bibfield  {journal} {\bibinfo  {journal}
  {Quarterly Journal of the Royal Meteorological Society}\ }\textbf {\bibinfo
  {volume} {114}},\ \bibinfo {pages} {1435} (\bibinfo {year}
  {1988})}\BibitemShut {NoStop}%
\bibitem [{\citenamefont {He}\ \emph {et~al.}(2009)\citenamefont {He},
  \citenamefont {Wu},\ and\ \citenamefont {Wang}}]{Heetal09}%
  \BibitemOpen
  \bibfield  {author} {\bibinfo {author} {\bibfnamefont {Z.}~\bibnamefont
  {He}}, \bibinfo {author} {\bibfnamefont {W.}~\bibnamefont {Wu}}, \ and\
  \bibinfo {author} {\bibfnamefont {S.}~\bibnamefont {Wang}},\ }\href {\doibase
  10.1061/(ASCE)HY.1943-7900.0000116} {\bibfield  {journal} {\bibinfo
  {journal} {Journal of Hydraulic Engineering}\ }\textbf {\bibinfo {volume}
  {135}},\ \bibinfo {pages} {1028} (\bibinfo {year} {2009})}\BibitemShut
  {NoStop}%
\bibitem [{\citenamefont {Wu}\ \emph {et~al.}(2012)\citenamefont {Wu},
  \citenamefont {Marsooli},\ and\ \citenamefont {He}}]{Wuetal12}%
  \BibitemOpen
  \bibfield  {author} {\bibinfo {author} {\bibfnamefont {W.}~\bibnamefont
  {Wu}}, \bibinfo {author} {\bibfnamefont {R.}~\bibnamefont {Marsooli}}, \ and\
  \bibinfo {author} {\bibfnamefont {Z.}~\bibnamefont {He}},\ }\href {\doibase
  10.1061/(ASCE)HY.1943-7900.0000116} {\bibfield  {journal} {\bibinfo
  {journal} {Journal of Hydraulic Engineering}\ }\textbf {\bibinfo {volume}
  {138}},\ \bibinfo {pages} {503} (\bibinfo {year} {2012})}\BibitemShut
  {NoStop}%
\bibitem [{\citenamefont {Duc}\ and\ \citenamefont {Rodi}(2008)}]{DucRodi08}%
  \BibitemOpen
  \bibfield  {author} {\bibinfo {author} {\bibfnamefont {B.~M.}\ \bibnamefont
  {Duc}}\ and\ \bibinfo {author} {\bibfnamefont {W.}~\bibnamefont {Rodi}},\
  }\href {\doibase 10.1061/(ASCE)0733-9429(2008)134:4(367)} {\bibfield
  {journal} {\bibinfo  {journal} {Journal of Hydraulic Engineering}\ }\textbf
  {\bibinfo {volume} {134}},\ \bibinfo {pages} {367} (\bibinfo {year}
  {2008})}\BibitemShut {NoStop}%
\bibitem [{\citenamefont {Cao}\ \emph {et~al.}(2011)\citenamefont {Cao},
  \citenamefont {Hu},\ and\ \citenamefont {Pender}}]{Caoetal11}%
  \BibitemOpen
  \bibfield  {author} {\bibinfo {author} {\bibfnamefont {Z.}~\bibnamefont
  {Cao}}, \bibinfo {author} {\bibfnamefont {P.}~\bibnamefont {Hu}}, \ and\
  \bibinfo {author} {\bibfnamefont {G.}~\bibnamefont {Pender}},\ }\href
  {\doibase 10.1061/(ASCE)HY.1943-7900.0000296} {\bibfield  {journal} {\bibinfo
   {journal} {Journal of Hydraulic Engineering}\ }\textbf {\bibinfo {volume}
  {137}},\ \bibinfo {pages} {267} (\bibinfo {year} {2011})}\BibitemShut
  {NoStop}%
\bibitem [{\citenamefont {Cao}\ \emph {et~al.}(2012)\citenamefont {Cao},
  \citenamefont {Li}, \citenamefont {Pender},\ and\ \citenamefont
  {Hu}}]{Caoetal12}%
  \BibitemOpen
  \bibfield  {author} {\bibinfo {author} {\bibfnamefont {Z.}~\bibnamefont
  {Cao}}, \bibinfo {author} {\bibfnamefont {Z.}~\bibnamefont {Li}}, \bibinfo
  {author} {\bibfnamefont {G.}~\bibnamefont {Pender}}, \ and\ \bibinfo {author}
  {\bibfnamefont {P.}~\bibnamefont {Hu}},\ }\href {\doibase
  10.1680/wama.10.00035} {\bibfield  {journal} {\bibinfo  {journal}
  {Proceedings of the ICE - Water Management}\ }\textbf {\bibinfo {volume}
  {165}},\ \bibinfo {pages} {193} (\bibinfo {year} {2012})}\BibitemShut
  {NoStop}%
\end{thebibliography}
%

\end{document}